\documentclass{article}
\usepackage{arxiv}

\usepackage[final]{fixme}

\usepackage{amsmath}
\usepackage{amssymb}
\usepackage{amsfonts}       

\usepackage[utf8]{inputenc} 
\usepackage[T1]{fontenc}    
\usepackage{url}            

\usepackage{multirow}
\usepackage{mathtools}
\usepackage{booktabs}       
\usepackage{caption}
\usepackage{subcaption}
\usepackage[page]{appendix}

\usepackage{graphicx}
\usepackage{tikz}
\usepackage[hidelinks]{hyperref}
\usepackage[gen]{eurosym}

\usepackage{nicefrac}       
\usepackage{microtype}      

\DeclareMathOperator*{\E}{\mathbb{E}}

\title{Targeting Customers under Response-Dependent Costs}

\author{
  Johannes Haupt
    \\
  School of Business and Economics\\
  Humboldt-Universit\"at zu Berlin\\
  \texttt{johannes.haupt@remerge.io} \\
   \And
  Stefan Lessmann\\
  School of Business and Economics\\
  Humboldt-Universit\"at zu Berlin\\
  \texttt{stefan.lessmann@hu-berlin.de} \\
}

\begin{document}
\maketitle

\begin{abstract}
This study provides a formal analysis of the customer targeting problem when the cost for a marketing action depends on the customer response and proposes a framework to estimate the decision variables for campaign profit optimization. 
Targeting a customer is profitable if the impact and associated profit of the marketing treatment are higher than its cost.
Despite the growing literature on uplift models to identify the strongest treatment-responders, no research has investigated optimal targeting when the costs of the treatment are unknown at the time of the targeting decision. Stochastic costs are ubiquitous in direct marketing and customer retention campaigns because marketing incentives are conditioned on a positive customer response. 
This study makes two contributions to the literature, which are evaluated on an e-commerce coupon targeting campaign. 
First, we formally analyze the targeting decision problem under response-dependent costs. Profit-optimal targeting requires an estimate of the treatment effect on the customer and an estimate of the customer response probability under treatment. The empirical results demonstrate that the consideration of treatment cost substantially increases campaign profit when used for customer targeting in combination with an estimate of the average or customer-level treatment effect.
Second, we propose a framework to jointly estimate the treatment effect and the response probability by combining methods for causal inference with a hurdle mixture model. The proposed causal hurdle model achieves competitive campaign profit while streamlining model building. \\
All code is available at \url{https://github.com/Humboldt-WI/response-dependent-costs}.
\end{abstract}


\section{Introduction}

The paper studies the targeting of marketing communication when the realized cost of an offer depends on the customer response. Conditional offers are common in digital marketing, which has become the primary channel for advertising and reaching out to customers. Market analysts estimate global spending on digital advertising to be 460.6 billion USD by 2024, compared to 335.5 billion USD in 2019 \cite{statista2020_digiadvertising}. Direct marketing has emphasized the criticality of addressing the right customers in the right way at the right time for decades \cite{Bult1995_optimal_selection}. Digital channels provide the data to act on this call. They facilitate tracing recipients' interaction with marketing communication and provide fine-grained behavioral data to optimize marketing policies \cite{Wedel16_marketing_analytics}. Revenues from digital advertising are a crucial source of revenue for many actors in the digital economy, which stresses the relevance of a suitable customer targeting methodology. Targeting is profitable if the positive impact of the marketing treatment on customer behavior, and the resulting revenue to the company from that behavior, is higher than the costs of the treatment. 
Previous studies focus on the prediction of expected revenue. We expand this perspective by emphasizing the prediction of expected costs and developing a profit-based targeting rule.

Predicting the expected profit of a marketing campaign requires estimating the change in customer behavior if the customer is targeted; known as conditional average treatment effect (CATE) \cite{varian16_causalinference} or uplift \cite{devriendt2018LiteratureSurveyExperimental}. 
Much recent work on causal machine learning (ML) develops scalable approaches for estimating the CATE from large, high-dimensional data sets \cite{athey2019GeneralizedRandomForests, Knaus_causal_ML_MC_study}. These advancements facilitate predicting heterogeneous responses to marketing treatment and, thereby, the revenue from targeted marketing.

Beyond revenues, a profitable targeting policy requires the consideration of treatment costs. However, prior work devotes little attention to the cost structures of customer targeting. For example, instead of considering costs as decision variables, several studies assume an external restriction on the number of customers to target and focus on identifying a corresponding number of most responsive clients \cite{ascarza2018RetentionFutilityTargeting}. 
Other studies develop profit-maximizing targeting policies but focus on settings in which the cost of the treatment is known at the time of the targeting decision \cite{hitsch2018HeterogeneousTreatmentEffects}; the costs of printing and shipping a catalog in the mail-order industry being an example. 

This paper elaborates on the cost structures of marketing treatments and their policy implications. First, the implementation of a marketing campaign entails \textit{fixed costs} for the design of the marketing action and the development of the targeting policy. While the fixed costs are an important strategic consideration, they do not affect operational targeting decisions for individual customers. To inform targeting decisions, we focus on variable costs associated with the marketing treatment and the communication channel. We distinguish between \textit{targeting-dependent} and \textit{response-dependent} variable costs. 

Targeting-dependent costs arise for the communication with the customer in the form of mail charges or call center fees and for the production of material treatments like catalogs \cite{hruschka2010ConsideringEndogeneityOptimal}. An important characteristic of targeting-dependent costs is that they arise when the targeting decision is made, independent of its success. 

Many marketing applications include costs that are uncertain when making the targeting decision because they are realized only if the customer accepts the marketing offer. These response-dependent costs emerge whenever a marketing incentive is conditional on customers performing some actions in response to receiving the treatment. Discounts are an example of conditional incentives and are used frequently in customer retention management \cite{ascarza2013JointModelUsage} and coupon targeting \cite{sahni2016TargetedDiscountOffers}. Because the treatment cost is conditional on the customer action, the uncertainty about the customer action translates into uncertainty about the realization of the incentive cost. The targeting decision must then be based on comparing the expected revenue to the expected cost of the marketing treatment, which is uncertain but can be estimated. 

Despite the prevalence of response-dependent costs in the industry and extensive research on customer retention and couponing, the literature has not analyzed targeting decisions under response-dependent costs as a policy problem and lacks modeling strategies to estimate both the treatment effect and customer choice. 
Striving to close this research gap, the paper makes three contributions. 
First, we provide a formal analysis of the targeting decision problem under different types of costs and exemplify common application settings. For the case of response-dependent costs, we show that a profit-maximizing targeting policy requires an estimate of the expected change in profit in the form of the CATE and an estimate of the customer response probability under treatment. More generally, our analysis provides decision rules for optimal targeting under different cost settings and identifies the parameters that marketers have to estimate from data. 

Second, we propose a modeling framework to implement the targeting decision rule. While focusing on campaigns that exhibit response-dependent costs, the framework is generic and encompasses other campaign settings as special cases. Our multi-model approach comprises joint estimation of the two key parameters of interest, the treatment effect on profit and the absolute response probability. 
We achieve this by integrating meta-learners for CATE estimation with an ML-based hurdle model. A benefit of the latter model is that it allows disentangling the treatment effect on the customer decision of \textit{whether} to respond (e.g., whether to redeem a coupon) and \textit{how} to respond (e.g., how much to spend after deciding to buy).

Last, the systematic evaluation of our framework makes an empirical contribution to the literature on coupon targeting. We employ data from an e-commerce setting in which digital coupons were issued to stimulate sales. E-coupons enjoy much popularity and were identified as the most important marketing trigger to complete a transaction by US consumers \cite{deal_seekers}. Our results quantify the effectiveness of a specific e-coupon campaign as a byproduct. More importantly, the observed results establish the suitability of framework components and suggest that the proposed analytical targeting policy yields higher profit than the empirical policies used in prior research. We also demonstrate the hurdle model to perform competitively to state-of-the-art ML methods for CATE estimation while providing additional insights into customers' reaction to a marketing treatment. This way, our results also contribute case-based evidence to the empirical literature on causal ML. 

\section{Literature Review}\label{literature_review}%
Prior work focuses on an empirical optimization of targeting decisions by predicting customer responses and treatment effects. Less attention has been paid to the cost structures of different campaigns and the decision-theoretic foundation of the targeting policy. 

\subsection{Empirical Optimization of the Targeting Policy}
Research on customer response behavior prediction advocates an empirical optimization of the targeting policy. Many studies use supervised ML for predicting customer-level entry probabilities of events of interest including purchase, churn, social media engagement, etc. The corresponding models are commonly assessed using the lift measure, the underlying idea of which is to rank-order candidate recipients of a campaign by their model-estimated response probabilities and to target a small fraction of top-ranked clients, depending on the available marketing budget \cite{DeCaigny_logit_leaf}. Some studies have proposed measures that capture business value more accurately than the lift measure  \cite{verbeke2012NewInsightsChurn, lessmann2019TargetingCustomersProfit}. However, basing targeting decisions on behavior predictions favors natural responders and ignores the causal effect of a treatment on customer responses \cite{ ascarza2018RetentionFutilityTargeting}. Therefore, recent research advocates estimating the treatment effect based on observed customer characteristics (i.e., CATE). 

One may categorize previous targeting models according to the marketing task they support, which depends on the scale of the outcome variable and the number of treatments. CATE estimators for a single treatment and binary outcomes prevail in the uplift modeling literature \cite{devriendt2018LiteratureSurveyExperimental}. They facilitate targeting campaigns aiming at conversion. CATE estimates capture how much a campaign raises a recipient's probability of performing the desired action (clicking, buying, liking, etc.). Campaigns aiming at revenue growth, on the other hand, require a CATE estimator for continuous outcomes, which predicts the incremental revenue due to targeting \cite{gubela2020ResponseTransformationProfit}. If a marketer has multiple options to solicit customers (over different channels, with different messages, etc.), uplift models for multiple treatments facilitate determining the most effective action to increase conversion \cite{Olaya_multi_treatment} or revenue \cite{gubela_multi_treatment_revenue_uplift}. 

Many studies adopt a methodological perspective and develop new ML-based algorithms for CATE estimation. Tree-based ensembles such as causal Bayesian additive regression trees \cite{hahn_causal_bart} or the generalized random forest \cite{athey2019GeneralizedRandomForests} are especially prominent. Echoing the interest in deep learning, architectures for estimating the CATE using neural networks have also been proposed \cite{Farrell_residual_nn}. Another stream of the literature proposes meta-learners, which facilitate CATE estimation using any supervised ML algorithm \cite{kunzel2019MetalearnersEstimatingHeterogeneous}. \cite{Knaus_causal_ML_MC_study} provide a comprehensive survey and empirical evaluation of corresponding approaches. 

Increasing sophistication in estimating the CATE has not altered the translation of model estimates into a targeting policy. Recent studies use the CATE as a ranking criterion, as opposed to estimated response probabilities, and recommend targeting a fraction of top-ranked clients \cite{hansotia2002IncrementalValueModeling, ascarza2018RetentionFutilityTargeting, gubela2019ConversionUpliftEcommerce}. 
The optimal proportion of the population to target can be approximated by comparing the group-wise average treatment effect for customers within each decile of the CATE estimates. A correct ranking of customers by their CATE implies that the average treatment effect in groups with high model estimates must be higher than in groups with low model estimates. The evaluation of the model's ability to rank customers by their expected treatment effect is in line with industry practice to target a small group of the most responsive customers but ignores the cost of targeting to determine the size of the campaign. An advantage of the empirical approach is that it remains feasible when the CATE estimates are a biased or badly calibrated estimate of the individual treatment effect (ITE) or when the profit and costs parameters of the campaign are unknown. However, when there is heterogeneity in response value or costs, ranking the customer by their expected treatment effect ignores variation in expected profit that is not due to variation in sensitivity to the treatment. Response-dependent costs imply such variation in expected cost even when the nominal cost of the treatment is constant. Therefore, under value or cost heterogeneity, empirical thresholding of the treatment effect will not result in an optimal targeting policy. 

\cite{lemmens2020ManagingChurnMaximize} offer the latest and most advanced instance of the empirical policy optimization paradigm. They estimate a score for the incremental net profit after targeting costs in a two-step procedure, which combines cost-sensitive learning with predicting a first-stage estimate of the CATE directly from customer characteristics \cite{hitsch2018HeterogeneousTreatmentEffects}. The first stage produces a CATE estimate using multiple models that are replaced by a single model in the second step. The approach integrates heterogeneous treatment costs into the customer scoring function, but requires a separate method to estimate the incremental net profit in the first stage and relies on empirical thresholding to determine an appropriate score above which to target customers. 
Compared to \cite{lemmens2020ManagingChurnMaximize}, we propose a method to estimate the expected incremental profit using a causal hurdle model. More importantly, we develop an analytical targeting policy to determine optimal campaign sizes and targeting decisions.

\subsection{Decision-Theoretic Optimization of the Targeting Policy}
Early work approaching customer targeting as a policy problem includes \cite{Bult1995_optimal_selection}, who formulate the classic decision rule of marginal costs equating marginal returns for a direct-marketing setting. Extending this idea by conditioning customer choices on targeting decisions,\cite{hansotia2002IncrementalValueModeling} show that targeting a customer is profitable when the incremental value of the marketing action is at least as high as its cost. 
\cite{hitsch2018HeterogeneousTreatmentEffects} further elaborate on targeting costs and propose a decision rule for targeting catalog mailings.
However, none of these studies considers response-dependent treatment costs.

\cite{neslin2006DefectionDetectionMeasuring} propose a profit function for customer retention campaigns. 
Next to targeting-dependent contact costs, their profit function includes a response-dependent cost for the incentive if a customer accepts the firm's retention offer but implicitly assumes a constant treatment effect\footnote{We discuss some interesting yet intricate connections between the profit function of \cite{neslin2006DefectionDetectionMeasuring} and our decision analysis in online \ref{sec:churn_profit_formula}, which also offers a generalization of the profit function.}. 
Later studies relax some of the assumptions of the profit function. 
\cite{verbraken2012NovelProfitMaximizing} introduce the expected maximum profit criterion for churn, which relaxes the assumptions of a fixed campaign size and a constant, a priori known probability of customers to accept the offer. \cite{devriendt2019WhyYouShould} further advance the profit criterion for uplift modeling, which facilitates accounting for heterogeneity in customer responses conditional on treatment. A limitation of their formulation is that it assumes the treatment effect to be strictly positive, which contradicts the categorization of customers into four stereotypes (e.g., do-not-disturbs, lost causes, sure things, and persuadables), which is instrumental to uplift modeling \cite{devriendt2018LiteratureSurveyExperimental}. Further, \cite{devriendt2019WhyYouShould} emphasize the evaluation of targeting models but do not provide guidance on how to use their profit function for model estimation.   

We conclude that analytical approaches for optimal targeting under response-dependent costs are limited to retention campaigns and rarely adopt a causal perspective, \cite{devriendt2019WhyYouShould} being the only exception. We contribute to the literature by providing a formal analysis of the general targeting decision problem, which encompasses retention campaigns as a special case. To facilitate optimal targeting, our proposition considers variation in treatment effects and variation in the expected realized cost over customers. We also avoid assuming positive treatment effects.

\section{Methodology}
In the following, we derive decision rules for customer targeting for campaign settings with different cost structures. Next, we develop a modeling framework to implement our policy recommendation. 
We focus on the binary decision setting, which prevails in prior work \cite{sahni2016TargetedDiscountOffers, gubela2020ResponseTransformationProfit, devriendt2018LiteratureSurveyExperimental}, and consider a marketer who decides whether to treat a customer with a given campaign.
However, prescribing a sound way to appraise the profitability of a treatment, our analysis is also relevant for multiple treatment settings, in which marketers aim at identifying the most suitable treatment out of a set of alternatives. Our modeling framework is directly applicable to such settings by decomposing the multi-treatment problem into a sequence of binary decisions \cite{Olaya_multi_treatment}.

\subsection{Optimal Decision Making in Customer Targeting}\label{response-dependent_costs}

The customer targeting decision problem is characterized by three components, 1) the value to the marketer conditional on the customer response, 2) the treatment cost conditional on targeting, and 3) the treatment cost conditional on the customer response. The existence of response-dependent costs differentiates retention and coupon campaigns from the cost setting discussed in previous studies \cite[e.g.]{hansotia2002IncrementalValueModeling}, which assume that all cost components are conditional on the targeting decision but independent of the customer response. 

Let $C_i \in {0,1}$ be a random variable indicating conversion of customer $i$, who is characterized by observed covariates $X_i\in\mathbb{R}^p$. We define conversion $C_i=1$ as an event with a positive impact on profit (e.g., a customer purchase). Let $V_i \in \mathbb{R}^+$ be the gross profit before targeting costs that is associated with a positive customer action. $V_i$ represents the customer lifetime value in churn prevention or the margin of a purchase in direct marketing and may vary across customers. For convenience, let $Y_i=C_i \cdot V_i$ be the observed profit of the targeting decision, excluding targeting cost. Note that $Y_i=V_i$ when $C_i=1$ and $Y_i=0$ otherwise. The probability of a positive response $p(C_i=1|X_i=x)$ and the expected response value $E[V_i|X_i=x]$ are unknown at the time of the marketing decision and need to be estimated given the customer characteristics. 

Recall that the variable costs split into two components, targeting-dependent and response-dependent costs. Let $\kappa$ be a targeting-dependent cost that is constant and independent of the customer characteristics. Targeting-dependent costs can be contact costs such as mail charges. Let $\delta$ be a response-dependent cost that applies if the customer responds positively after receiving the marketing treatment.  The response-dependent cost can be associated with a marketing incentive that is conditioned on a positive customer response, for example, a voucher for free shipping for the current purchase process. The expected response-dependent cost at the time of targeting depends on the probability that the customer will accept the offer. Besides, response-dependent costs may depend on the value of the response. When the marketing treatment is a relative discount, for example, in the form of a 10\% discount on the current purchase, the nominal discount depends on the completion of the purchase and the purchase amount. The expected offer cost then depends on the probability of positive customer response and the value of the response. If a customer is not targeted by the campaign, then no variable costs occur and $\delta=\kappa=0$. 

Table \ref{tab:decision_summary} summarizes decision problems in target marketing by outlining their cost structure and anticipates the results of the decision analysis. The decision problems vary in the existence of treatment- and response-dependent costs, the type of response-dependent incentive, and assumptions about the treatment effect on response probability and value. We see that targeting-dependent costs apply to one stream of research with applications in catalog marketing \cite{hitsch2018HeterogeneousTreatmentEffects} and online banner advertising \cite{diemert2018LargeScaleBenchmark}. The proposed decision framework applies under any combination of variable costs and is crucial whenever there are response-dependent costs. We further differentiate the response-dependent costs into offers with a fixed value, e.g. retention campaigns with a discount upon contract renewal \cite{devriendt2019WhyYouShould}, and offers with a value equal to a percentage of the response value, e.g. coupons \cite{gubela2019ConversionUpliftEcommerce}. The last column of Table \ref{tab:decision_summary} provides the decision variables per setting. These are the variables required to calculate the expected profit of the targeting decision as a result of our analysis. The set of decision variables may simplify when assuming no treatment effect on the value given conversion. Below, we discuss this assumption as a special case.

We associate digital channels with treatment-dependent costs of zero in Table~\ref{tab:decision_summary}. One may challenge this choice, referring to, for example, the sizeable investments of firms into email marketing or the costs associated with banner ads on 3rd party websites using real-time bidding systems. Our justification for assuming zero treatment costs is twofold. First, we consider micro-level decisions on an individual customer level. The corresponding treatment costs (e.g., the cost of sending one email), are typically negligible in digital channels. Second, the novelty of the paper lies in the handling of response-dependent costs, which have eluded research. By definition, treatment costs are independent of customer responses. They enter the decision rules, which we derive below, as constants. Therefore, it is straightforward to consider treatment costs whenever needed. 

\begin{table}[ht]
\centering
\small
\caption[Decision problems in customer targeting and their decision variables]{Decision problems in customer targeting and their decision variables}
\label{tab:decision_summary}
\begin{tabular}{lcccccl}
\toprule
   & \multicolumn{3}{c}{Cost} & &  &  \\
 \cmidrule(lr){2-4}
& Treat.-Depend. & \multicolumn{2}{c}{Resp.-Depend.}& \multicolumn{2}{c}{Treat. Effect on}\\
\cmidrule(l{2pt}r{2pt}){3-4}\cmidrule(l{2pt}r{2pt}){5-6}
  Application Example        &    & Fixed  & Percent. &    Decision     & Value    &     Decision Variables         \\
          \midrule
   \textit{Advertisement}\\
Letter and Present\textsuperscript{1}   & yes        & no       & no         & yes    & no   & $p_i(1)-p_i(0), V_i$   \\
Online Ad Banner\textsuperscript{2}    & no         & no       & no         & yes    & no  & $p_i(1)-p_i(0)$ \\

Catalog\textsuperscript{3}   & yes        & no       & no         & yes    & yes  & $Y_i(1)-Y_i(0)$       \\
Online Ad Banner    & no         & no       & no         & yes    & yes  & $Y_i(1)-Y_i(0)$ \\
\\
\textit{Discount Coupon}\\
Print Retention Offer\textsuperscript{4} & yes         & yes       & no         & yes    & yes  & $Y_i(1)-Y_i(0), p_i(1)$    \\
Online Fixed Value  & no         & yes       & no         & yes    & yes  & $Y_i(1)-Y_i(0), p_i(1)$ \\
\\
Print Discount Coupon     & yes         & no       & yes         & yes    & yes  & $Y_i(1)-Y_i(0), p_i(1), V_i(1)$    \\
Online Percent. Coupon\textsuperscript{5}  & no         & no        & yes        & yes    & yes  &$Y_i(1)-Y_i(0), p_i(1), V_i(1)$\\
Online Percent. Coupon & no         & no        & yes        & yes    & no   & $p_i(1), p_i(0)$ \\
\bottomrule
\multicolumn{7}{l}{
 \textsuperscript{1}\cite{ascarza2018RetentionFutilityTargeting} \,
 \textsuperscript{2}\cite{diemert2018LargeScaleBenchmark} \,
 \textsuperscript{3}\cite{hitsch2018HeterogeneousTreatmentEffects} \,
 \textsuperscript{4}\cite{devriendt2019WhyYouShould} \,
 \textsuperscript{5}\cite{gubela2019ConversionUpliftEcommerce}}
\end{tabular}
\end{table}

%

Let $T_i$ be a variable indicating whether a given marketing treatment was applied to customer $i$. We consider a single treatment setting and assume $T_i \in {0,1}$, where $T=1$ indicates that the customer is targeted, the treatment condition, and $T=0$ indicates that she is not; the control condition. The following analysis is easily extended to more than one treatment by considering multiple binary comparisons. The treatment is designed to increase the conversion probability of the customer or her value given conversion, or both. 
Following the Neyman-Rubin potential outcome model, we denote the potential outcomes under treatment by $\cdot(0)$ and $\cdot(1)$. For example, $C_i(1)$ and $C_i(0)$ denote the conversion outcome if customer $i$ is targeted and not targeted, respectively. The individual treatment effect (ITE) on profit is then $\tau_i = Y_i(1) - Y_i(0) = C_i(1)V_i(1) - C_i(0)V_i(0)$. We further distinguish between the ITE on response probability $\tau^C_i = C_i(1) - C_i(0)$ and the ITE on response value $\tau^V_i = V_i(1) - V_i(0)$.

We now begin our analysis of the targeting decision problem. The profit $\pi_i$ for an individual in the marketing campaign including treatment costs is
\begin{align*}
\pi_i = 
    \begin{cases}
    C_i(0) V_i(0) & \text{if } T_i =0\\
    C_i(1)   V_i(1) - C_i(1) \delta - \kappa & \text{if } T_i=1
    \end{cases}
\end{align*}

The general decision problem of whether to target a specific customer under response-dependent costs can then be posed as
\begin{align}
p_i(1) (V_i(1) - \delta) - \kappa > p_i(0) \cdot V_i(0) ,
\end{align}

where we use $p_i$ as a shorthand for $p(C_i=1|X_i=x)$. Note how variable costs affect the campaign profit. The target-dependent costs $\kappa$ are realized before the customer makes a decision and are, therefore, independent of the customer action. The response-dependent costs $\delta$ are realized only when a positive response takes place. Solving the inequality for the treatment effect yields
\begin{align}
p_i(1) V_i(1) - p_i(0) V_i(0) > p_i(1) \delta + \kappa \label{eq:decision_rule}
\end{align}

The optimal decision naturally depends on the individual treatment effect on the profit on the left side of the inequality. However, it also depends on the probability of a positive customer response under treatment as a mitigating factor on the offer cost. Intuitively, the absolute offer costs are a promise from the firm and must be discounted by the chance that the customer will demand fulfillment. 
The response-dependent costs are not incurred if the customer refuses the offer.
Therefore, the customer targeting decision under response-dependent costs differs from the case where $\delta=0$ because the costs are now stochastic 
at the point of the targeting decision. 

The optimization of expected profit underlying Eq. \ref{eq:decision_rule} implies that, when faced with two customers with an identical CATE, it is more profitable to target the customer who is less likely to accept the marketing offer. Targeting customers with a high response probability after treatment not only disregards the causal effect of the treatment, as prior work has pointed out \cite{ascarza2018RetentionFutilityTargeting}, but increases the cost of campaigns by targeting customers with high expected response-dependent costs. To clarify the intuition behind this result, consider the treatment of a customer as an investment with probabilistic costs. If the payout of two investments is identical, a rational agent prefers the investment that has lower expected costs. 
This result suggests that when there is little or no treatment heterogeneity, meaning that the payout of the treatment is identical between customers, it is profitable to target customers with a lower rather than higher probability to respond.

Eq. \ref{eq:decision_rule} reveals that decision settings with response-dependent costs require an estimate of the treatment effect $p_i(1) V_i(1) - p_i(0) V_i(0)$ and an estimate of the response probability $p_i(1)$. This result is surprising because prior literature emphasizes uplift models, which provide an estimate of the treatment effect, as a direct replacement of response models, which provide an estimate of the conversion probability. 
We show that decisions under response-dependent costs require both, a model of the treatment and a model of the conversion probability under treatment. 

In practice, a positive expected profit may not result in an optimal policy under strategic considerations. Actual targeting campaigns are regularly evaluated by their return on advertising spend (ROAS). The ROAS is defined as the ratio of campaign profit over campaign costs. 
Being a metric of advertising efficiency, the ROAS does not consider campaign size. While it is generally not profit-optimal to maximize efficiency at the cost of targeting fewer customers, a minimum ROAS is often required in practice to satisfy management goals and allocate resources efficiently between marketing channels or campaigns. A side result of our analysis is that the proposed decision rule can be used to set targeting thresholds to reflect a minimum ROAS as
\begin{align}
\frac{p_i(1) V_i(1) - p_i(0) V_i(0)}{p_i(1) \cdot \delta + \kappa  } \geq \text{Target ROAS}             \nonumber  
\end{align}
We proceed by discussing two special cases that arise in digital applications. 
First, assume that $\kappa=0$. Targeting-dependent (contact) costs of zero are plausible when communicating with customers via digital channels. For example, the costs for email targeting and banner campaigns on a company's websites arise in the form of a fixed cost into infrastructure (e.g., content management systems and content production). These costs are irrelevant for operational targeting decisions in the short run. Rearranging Eq. \ref{eq:decision_rule}, the targeting rule becomes 
\begin{align}
\label{eq:kappa=0}
p_i(1) (V_i(1)-\delta)&> p_i(0) V_i(0)
\end{align}
%

Assume additionally that offer costs depend linearly on the response value, i.e. $\delta_{i} = \eta V_i(1) $. This assumption corresponds to discount coupons that reduce the checkout amount by a fixed percentage and other forms of dynamic pricing. Percentage discount coupons are frequently used in digital marketing as a transparent means to differentiate incentives according to the value of customers and as an incentive that encourages higher spending. Substituting $\delta$ in Eq. \ref{eq:decision_rule} with $\delta_i$, we find that the decision rule for such discount offers also requires an estimate of the expected response value under treatment
\begin{align}
p_i(1) V_i(1) - p_i(0) V_i(0)  &> p_i(1) \cdot \eta \cdot V_i(1) \label{eq:discount_coupon}  
\end{align}
%
%
Second, there exists a special case of the decision problem in Eq. \ref{eq:discount_coupon} that requires no estimate of the purchase value. Assume that the treatment affects the conversion probability but not the response value, i.e. $V_i(1)=V_i(0)=V_i$. Then, equation \ref{eq:discount_coupon} reduces to
\begin{align}
(p_i(1) - p_i(0)) \cdot V_i > p_{i}(1) \cdot \eta \cdot V_i  \Rightarrow  p_{i}(1) > \frac{p_{i}(0) }{1- \eta} \label{eq:discount_simple}
\end{align}

%
Under the combined assumptions of a percentage discount with no fixed contact cost and no effect on conversion value, the decision rule does not depend on the individual purchase value. Intuitively, a negligible communication cost removes the need to make up for the cost of customer targeting. Further making the coupon cost dependent on the response value automatically adjusts the cost to decrease with smaller response values and vice versa. In practice, this setting requires estimation of the purchase probabilities with and without treatment. 

The two special cases imply that the cost structure, which is determined by the infrastructure of the campaign and the design of the treatment, can increase or reduce the complexity of the decision problem. In general, when the cost of the treatment is conditioned on additional variables, then the estimation of these variables is relevant for the decision problem. We can see that percentage discounts on the purchase value introduce an estimate of the purchase value under treatment into the decision (Eq. \ref{eq:discount_coupon}). Similar arguments can be made for more specialized coupon designs like a minimum purchase value or a staggered discount increasing with purchase value. The second case shows that specific cost structures may simplify the decision problem. Under the additional assumption of no treatment effect on value, the targeting decision reduces to the estimation of the probabilities of purchase with and without treatment in Eq. \ref{eq:discount_simple}. 

The decision framework generalizes the decision settings discussed in prior work such as catalog marketing \cite{hansotia2002IncrementalValueModeling,hitsch2018HeterogeneousTreatmentEffects}. They consider campaigns with treatment costs but no response-dependent costs. Assuming $\delta=0$, the treatment effect on profit $Y_i$ is sufficient for the targeting decision, which reduces to
\begin{align}
p_i(1) V_i(1) - p_i(0) V_i(0) > \kappa \label{eq:catalog_decision}
\end{align}

We see immediately that an estimate of the treatment effect on the profit  
is a sufficient decision criterion under the conditions of Eq. \ref{eq:catalog_decision}. If we assume no treatment effect on the value such that $V_i(1)=V_i(0)=V_i$ and assume $V_i$ to be known or modeled independently, we recover
\begin{align}
p_i(1) - p_i(0) > \frac{\kappa}{V_i}, \label{eq:discount_decision}
\end{align}
where the focus lies on the estimation of the treatment effect on the customer response. This recovers the estimation problem addressed by many previous studies on uplift modeling although the corresponding papers rarely discuss the dependency on the response value or other assumptions concerning the campaign setting, with \cite[fn. 27]{ascarza2018RetentionFutilityTargeting} being an exception.

In summary, profit-based targeting involves additional decision variables beyond the treatment effect. Table \ref{tab:decision_summary} systematizes these decision variables and emphasizes that they depend on the cost structure of the marketing treatment.
For variable costs with a fixed value, the purchase probability under treatment determines the cost as in Eq. \ref{eq:decision_rule}. For treatment with a value relative to the purchase value, the purchase probability under treatment and the purchase value under treatment jointly determine the effective treatment cost as in Eq. \ref{eq:discount_coupon}. Both cost structures are common in direct marketing. Below, we propose a model specification to estimate the cost-related decision variables $p_i(1)$ and $V_i(1)$ within the model of the treatment effect $Y_i(1)-Y_i(0)$.

\subsection{Causal Hurdle Models}\label{causal_hurdle_models}
The targeting decision under response-dependent costs (Eq. \ref{eq:decision_rule}) requires estimates of the treatment effect $Y_i(1) - Y_i(0)$, the response probability under treatment $p_i(1)$, and the response value under treatment $V_i(1)$. 
Estimating each decision variable using one distinct model requires two models for estimating $p_i(1)$ and $V_i(1)$, respectively, in addition to the model for estimating the CATE $\hat{\tau}_i$, which requires one to four (sub-)models depending on the specific CATE estimation strategy. Hereafter, we call this approach the \textit{distinct modeling} approach.

We propose a framework that avoids building distinct models for the CATE and the customer response under treatment by simultaneously estimating the treatment effect on expected profit, the purchase probability, and the purchase value. The proposed framework exploits
the decomposition of the expected profit according to $Y_i=C_i \cdot V_i$ and collects estimates for $p_i(1)$  and $V_i(1)$ from the treatment effect model.
To achieve this, we estimate $V_i(1)$ and $p_i(1)$ jointly within the profit model by modeling the observed profit from customer $Y_i$ as a two-stage hurdle structure.

Hurdle models are mixture models over two distributions, one of which has a point mass at zero. 
They are suitable for decision processes that involve a binary decision of whether to act, the hurdle, and a conditional decision on the value associated with acting. Hurdle models assume that the occurrence of zeros is entirely driven by a first-stage process. The second stage value is zero when the first stage decision is a negative response and strictly positive when the first stage decision is a positive response \cite{donkers2006DerivingTargetSelection}.  

The hurdle model allows us to decompose the estimation of the profit $Y_i$ into the estimation of response $C_i$ and response value $V_i$. The probability mass function of the hurdle model is
\begin{align}
\label{eq:hurdle_model}
\Pr(Y_i| X_i=x) = 
\begin{cases}
    (1- \Pr(C_i=1|X_i = x))          \cdot  0                                             & \text{if } Y_i = 0   \\
          \Pr(C_i=1|X_i  =x )         \cdot   \Pr(V_i =v| X_i=x)  & \text{if } Y_i > 0 \\
\end{cases}
\end{align}

where $\Pr(C_i = 1 | X_i=x)$ and $\Pr(V_i =v |X_i=x)$ are models of the binary customer response and its value, respectively. If a customer responds, she decides on her spending behavior in the second stage, which determines the response value to the firm. The profit is zero if a customer chooses not to respond and strictly positive otherwise. 
Observing the spending only for customers who complete their purchase introduces selection bias. Customers who are less likely to convert are underrepresented in the data used to predict the basket value. Selection bias problems have received much attention and may be addressed using Heckman's correction \cite{Heckman1976}. In this paper, we estimate $\Pr(V_i =v|X_i=x)$ from the subset of customers where $C_i=1$ and weight the observed purchases by their estimated response probability to correct for selection bias. 

Two properties make the hurdle model specification suitable for modeling customer choice \cite{vandiepen2009DynamicCompetitiveEffects}.
First, the separation of the purchase and the value decision facilitates the estimation and interpretation of each model. For treatment estimation, separating the effect on response probability and response value provides a more nuanced understanding of marketing effectiveness and may help improve the treatment. The model structure also accommodates differences in the relevance of available covariates for each decision step \cite{donkers2006DerivingTargetSelection}. Second, the models for the prediction of response probability and value can be estimated separately when the purchase incidence is observed and we assume independent error terms \cite[p.545]{cameron2005MicroeconometricsMethodsApplications}. 
This property provides flexibility in specifying the component models. In the empirical study, we implement the response and value model using tree-based ensemble learners, which allows us to benefit from the scalability and expressive power of ML algorithms. Similar arguments have motivated the use of ML for estimating nuisance functions in causal models \cite{Knaus_causal_ML_MC_study}. Encouraging results from ML-based mixture models with hurdle structure have been observed in the scope of credit risk analytics \cite{Jiang_rf_hurdle_model} and support the proposed approach.

It remains to integrate the hurdle model into a framework for causal inference. 
Under the assumptions of the potential outcome framework, i.e. unconfoundedness, overlap, and stable unit treatment value, the CATE can be expressed as the difference between the outcome $Y_i$ conditional on treatment assignment $T_i$ and covariates $X_i$,
\begin{align}
\label{eq:difference_in_response}
\tau(x) = \E[Y_i(1)-Y_i(0)|X_i=x] & = \E[Y_i|X_i=x,T_i=1] - \E[Y_i|X_i=x,T_i=0].
\end{align}

We propose a hurdle model for estimating the conditional expectations in Eq. \ref{eq:difference_in_response}. 
\begin{flalign}
\label{eq:causal_hurdle_model}
\Pr(Y_i| X_i=x, T_i=t) = 
\begin{cases}
    1- \Pr(C_i=1|X_i=x, T_i=t)                                                   & \text{if } Y_i=0   \\
          \Pr(C_i=1|X_i=x, T_i=t)             \cdot   \Pr(V_i=v | X_i=x, T_i=t)  & \text{if } Y_i > 0 \\
\end{cases}&%
\end{flalign}

where the conversion probability and the purchase value conditional on conversion depend on the treatment assignment $T_i$ and the covariates $X_i$. 

Following the definition of the above hurdle model and under the assumptions of the potential outcome framework, we specify our causal hurdle model as
\begin{alignat}{2}
    \hat{\tau}(x)  & = \hspace{0.5em} &&\hat{y_i}(X_i,1) - \hat{y_i}(X_i,0) \nonumber \\ 
    &= &&p(C_i=1|X_i=x, T_i=1) \cdot \E[V_i|X_i=x,T_i=1]  \nonumber \\
    &  &&-p(C_i=1|X_i=x, T_i=0) \cdot \E[V_i|X_i=x,T_i=0]. \label{eq:causal_hurdle_effect}
\end{alignat}

Estimating the treatment effect on response probability and response value separately has an additional advantage if we expect heterogeneity of effect direction and size on customer value and response probability. This is the case if individual customers react differently to the same offer, for example, purchase their basket with higher probability or put additional products into their basket in response to receiving the treatment. Further, we expect the treatment effect on probability and value to be closely connected to the design of the marketing action. Under strong heterogeneity, we expect some customers to react to the marketing treatment by increasing the response value, e.g. putting more products into their basket, while becoming more reluctant to respond at the higher value, e.g. abandon a high-value shopping basket. Explicit estimates of the disentangled treatment effects are then relevant for treatment selection and design. 

\begin{figure}[ht]
    \centering
    \captionsetup{width=0.75\textwidth}
    \includegraphics[width=0.8\textwidth]{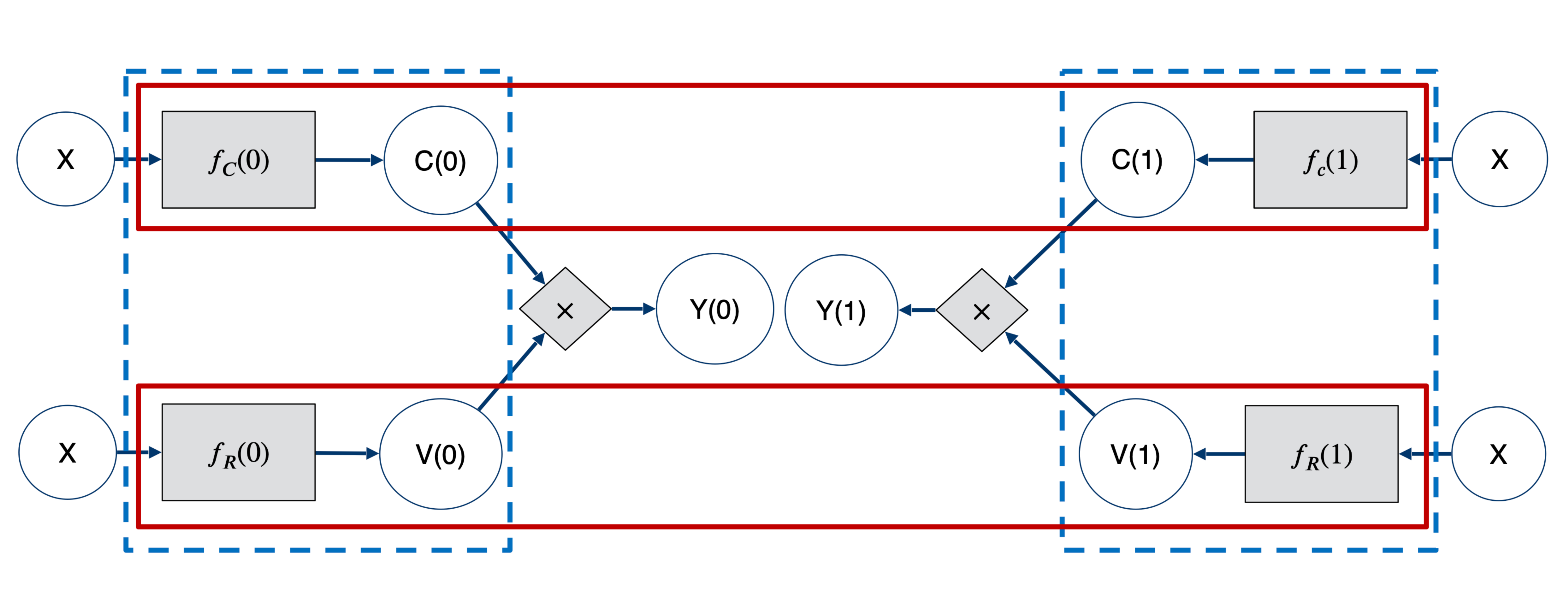} 
    \caption[Causal hurdle model structure]{Causal hurdle model structure. Frames indicate the two proposed estimation strategies: two hurdle models (dashed blue) or two causal single models (solid red)}
    \label{fig:causal_hurdle_structure}
\end{figure}

The formulation in Eq. \ref{eq:causal_hurdle_model} does not restrict the specific method of causal inference. Figure \ref{fig:causal_hurdle_structure} visualizes the general structure of causal hurdle models and makes the estimation targets explicit. We see that one strategy to estimate all relevant decision variables is to estimate four separate models, i.e. one model each for purchase probability and purchase value times one model each for the treatment and control group. This \textit{two-model hurdle model} is equivalent to combining the two-model approach for CATE estimation with two hurdle models for which the choice and value components are estimated separately \cite[p.545]{cameron2005MicroeconometricsMethodsApplications}.

It is possible to simplify the estimation by estimating more than one decision variable jointly. 
Eq. \ref{eq:difference_in_response} is the starting point for two approaches to integrating a hurdle model structure into treatment effect models. Figure \ref{fig:causal_hurdle_structure} visualizes the proposed methods to reduce the number of separate models by joint estimation of variables horizontally (solid red), over treatment and control group, or vertically (dashed blue), over purchase probability and value.

 The \textit{single-model hurdle model} combines the single model approach for causal inference with a two-stage estimation procedure for the hurdle model. A general model for the conditional profit with or without treatment takes the form $y=f(x,t)$ and predicts the return given the covariates $X_i$ and treatment assignment $T_i$. Despite its simplicity, the single model approach provides competitive CATE estimates for sufficiently flexible specifications of $f(\cdot)$  \cite{kunzel2019MetalearnersEstimatingHeterogeneous}. A flexible parametrization $f(\cdot)$ allows modeling the conditional average treatment effect through the interaction between $T_i$ and covariates $X_i$ within the model \cite{lemmens2020ManagingChurnMaximize}.
 The single model hurdle model includes the treatment variable as a covariate into the model and estimates one model for the response probability and one model for the customer value each jointly over the control and treatment group (Figure \ref{fig:causal_hurdle_structure}, solid red).

\section{Data and Experimental Design}\label{experimental_design}
We evaluate the proposed methodology in the context of targeting e-coupons.
Extensive use of coupons as a marketing instrument motivates the choice of this context. For example, US retailers distributed 256.5 billion coupons for consumer packaged goods in 2018, and consumers redeemed over 1.7 billion coupons with a combined face value of 2.7 billion USD \cite{NCH_coupon_facts}. For e-coupons, response-dependent costs in the form of the coupon value also dominate targeting-dependent costs (e.g., displaying a banner or sending an e-mail), which further motivates the choice of e-coupon targeting for the empirical study. 

The data comes from a German online fashion retailer 
and comprises 119,245 anonymized website visitors in the form of 61 variables collected through the retailer's shop system. 
The covariates describe the user and customer journey at the fifth page view of each customer's current visit when the retailer makes the decision whether to display a coupon to the user. The data contains aggregate information about previous visits to the website (where available), information on the current visit at each webpage view, and the outcome of the visit, including if a purchase was made and the value of that purchase. The covariates capture information on the user history (e.g., the number of previous visits), the behavior on the website (e.g., the number of clicks), and the current shopping basket. \cite{baumann2019PricePrivacyEvaluation} discuss the covariates in detail.
 
Out of all visits, 8,452 visits resulted in a purchase, which implies a conversion rate of 7\% conditional on a session length of at least five page views.
The value of purchases is distributed around a median of \EUR{73} with 90\% of purchases between \EUR{11.59} and \EUR{210} and a maximum value of \EUR{1035}. The covariates of the observations with the highest 1\% of purchase values do not indicate that these high basket values deviate systematically from purchases with a lower value or are caused by an error in the logging system during data collection. Therefore, we refrain from applying outlier corrections to observed purchase values or filtering visits with large purchases. 

The campaign of the e-tailor offered visitors an e-coupon through a banner, which stated the discount offer (\EUR{10} on purchases) and a coupon code to be entered upon checkout. The banner was shown repeatedly on subsequent page views to ensure that the customer is aware of the offer. 
The presence of the coupon banner might prohibit the e-tailor to display other banners to the visitor and may, therefore, carry an opportunity cost. While our framework supports non-zero targeting-dependent costs, we lack information to quantify their value (e.g., opportunity costs) and thus assume $\kappa=0$. The decision rule for campaign profit maximization is then given by Eq. \ref{eq:kappa=0} and advocates offering a coupon to the visitors whose incremental margin if targeted is larger than the expected cost of the coupon conditional on the customer's purchase probability.

To evaluate the proposed targeting policy and modeling framework, we simulate the treatment effect on customer spending on the actual customer data in an empirical Monte Carlo study. Simulation is suitable because implementing the profit-maximizing policy involves estimating the ITE of a coupon, the true value of which cannot be observed in practice. Therefore, the goal of the empirical Monte Carlo approach is to evaluate the proposed targeting framework on realistic customer data while controlling the ITE to facilitate evaluation in comparison to the ground truth. Anchoring the simulation in the observed data avoids oversimplifying the data generating process by assuming simple relations among the covariates and between the covariates and the outcome \cite{Knaus_causal_ML_MC_study, nie2017QuasiOracleEstimationHeterogeneous}. Further, to simulate realistically complex customer responses to the marketing incentive, we consider a process that depends nonlinearly on a subset of the observed covariates and define the range of the simulated ITE on response probability and value based on recent work of \cite{hitsch2018HeterogeneousTreatmentEffects}. We elaborate on the design of the simulation study in the online appendix (see Appendix \ref{appendix:te_simulation}).

We compare several options to instantiate the proposed modeling framework, which differ in the estimation strategies for the treatment effect and the response probability. Table \ref{tab:experiment_summary} summarizes the combinations considered in the study. 
We distinguish causal hurdle models and distinct modeling approaches. The former estimate the response probability and the response value given a positive response, with and without treatment. Among the causal hurdle models, we estimate the CATE using the single- and two-model approach \cite{kunzel2019MetalearnersEstimatingHeterogeneous}. Distinct modeling approaches estimate the treatment effect on profit in one stage but require a separate model to estimate the conversion probability under treatment. To estimate the CATE on profit, we again consider a single- and two-model approach and augment these with two state-of-the-art approaches that estimate the treatment effect directly from the covariates. The first of these is the Accelerated Bayesian Causal Forest (XBCF) \cite{hahn_causal_bart, BCF}. XBCF models the outcome as the sum of the expected outcome and the treatment effect, both of which are modeled as two Bayesian Additive Regression Trees (BART) and estimated jointly.
The second approach is the doubly-robust outcome transformation (DR). Under the DR approach, the treatment effect is estimated using a single model on a transformation of the profit 
\begin{align*}
&Y^{DR}_i = \mu_1 - \mu_0 + \frac{T_i(Y_i-\mu_1)}{p(T_i=1|X_i=x)} - \frac{(1-T_i)(Y_i-\mu_0)}{1-p(T_i=1|X_i=x)}  \\[1.5\jot]
&\text{with } \quad \mu_1 = E[Y_i|X_i=x, T_i=1] \quad \text{ and } \quad \mu_0 = E[Y_i|X_i=x, T_i=0]
\end{align*}
Three auxiliary models provide estimates for the expected profit in the treatment and control group, $E[Y_i|X_i=x, T_i=1]$ and $E[Y_i|X_i=x, T_i=0]$, and the probability to receive treatment, $p(T_i=1|X_i=x)$. To avoid instability when tuning the auxiliary models, we select linear regression to estimate the expected profit and logistic regression to estimate the probability to receive treatment.

With the exception of XBCF, the modeling approaches support any supervised ML algorithm for predicting customer response and value. However, including the treatment variable into the model, the singe-model approach requires a sufficiently flexible learner to capture interaction effects between the treatment indicator and covariates. In this study, we use gradient-boosted tree (GBT) ensembles as the underlying learning algorithm.%
We use five-fold cross-validation to compare different modeling approaches and perform an additional ten-fold (inner) cross-validation for tuning GBT  hyperparameters. To that end, we minimize the transformed outcome loss (TOL) by means of grid search over candidate hyperparameter settings. 

\begin{table}[ht]
\centering
\caption{Summary of model specifications considered in the experiment}
\label{tab:experiment_summary}
\begin{tabular}{llllc}
\toprule
          \multicolumn{4}{c}{Architecture}  & Number \\
\cmidrule(lr){1-4}
 Stages & CATE Model & Conversion Model  & Estimator & of Models \\
\midrule
\textit{Causal Hurdle Models}\\
  Hurdle &  single-model &               - &                  GBT  & 2\\
  Hurdle &     two-model &               - &                  GBT  & 4\\
  \textit{Distinct Modeling Approaches}\\
 One-Stage&  single-model &            Separate &                  GBT & 2  \\
One-Stage &     two-model &            Separate &                  GBT  & 3 \\
  One-Stage &            DR &            Separate &                  GBT & 5\\
    One-Stage &            BART &            Separate &                  GBT & 2\\
\bottomrule
\end{tabular}
\end{table}

\section{Empirical Results}\label{empirical_results}
The decision analysis defines a targeting policy as a combination of an estimate of the treatment effect and an estimate of the treatment cost, which depends on the conversion probability under treatment. The evaluation concentrates on framework components for estimating decision variables and their economic implications. We first assess estimates of the conversion probability. Next, we evaluate the precision of CATE estimates. Finally, we combine all framework components and test if our analytical targeting policy yields higher returns than a reference policy. 

\subsection{Profit Implications of Individual Cost Estimates} \label{sec:individual_cost_estimates}
We begin with the prediction of the conversion probability under treatment, which determines the expected cost of targeting. 
Table \ref{tab:conversion_profit} reports the profit and the fraction of customers treated for campaigns under the policy of Eq. \ref{eq:kappa=0}. 
The analysis includes several model-based approaches to estimate the conversion rate and, for comparison, the expected conversion rate in the population.
To calculate campaign profit, the conversion estimates are combined with two estimation procedures of the treatment effect. 
The treatment effect for each customer is either fixed as the average treatment effect (ATE) over all customers in the training data or presumed to be an \textit{oracle} estimate of the true ITE predicted by the model used in simulating the treatment effect.
The ATE policy makes the simplifying assumption that there exists no heterogeneity in treatment effects and is equivalent to the constant acceptance rate of the treatment assumed in prior studies on customer churn \cite[e.g.]{verbraken2012NovelProfitMaximizing}. Beyond the comparison to prior research, the ATE policy provides an estimate of the profit implication of cost-based targeting alone. The campaign profit under the simulated ITE provides an upper bound on campaign profit that would be achievable by adding estimation of individual-level cost to a treatment effect model with optimal performance. 

\begin{table}[htb]
\centering
\caption{Policy profit for the conversion models evaluated under selected treatment effect estimation methods}
\label{tab:conversion_profit}
\begin{tabular}{llllrr}
\toprule
&\multicolumn{3}{c}{Architecture} & \multicolumn{2}{c}{Campaign} \\
\cmidrule(lr){2-4}\cmidrule(lr){5-6}
Policy & CATE Model & Conversion Model & Estimator &  Profit & FT  \\
\midrule
Baseline &      -  &        -   &      -  &   50,668 &          0.00 \\
\addlinespace[1ex] 
   Analytical &      ATE &      -                 &         Conversion Rate &  55,055 &          1.00 \\
   Analytical &     ATE &            Single-Model &                  GBT &   55,289 &          0.72 \\
   Analytical &     ATE &               Two-Model/Distinct &                  GBT &  54,912 &          0.74 \\
   \addlinespace[1ex] 
   Analytical &          Oracle &    -       &         Conversion Rate &   60,125 &          0.56 \\
   Analytical &  Oracle &            Single-Model &                  GBT &  61,604 &          0.57 \\
   Analytical &  Oracle &               Two-Model/Distinct &                  GBT  &  61,504 &          0.57   \\
\bottomrule
\multicolumn{6}{l}{\small FT: fraction of customers targeted.} \\
\end{tabular}
\end{table}

Table \ref{tab:conversion_profit} shows that the baseline policy, under which no customer is targeted, yields a profit of \EUR{50,668}. This profit results from the natural probability in the customer population to complete a purchase, which we hope to increase with the marketing campaign. Next, consider the average treatment effect of the population and the average conversion rate of the population given treatment. The analytical policy recommends targeting all customers given the positive expected average return. The treatment rate of 100\% results in a profit of \EUR{55,055}. 
The fact that targeting all customers is more profitable than targeting no customers follows from the cost structure of coupon targeting (e.g., assuming zero communication costs). Due to the low organic conversion rates in e-commerce, the expected cost of the offer itself is also low. Given low costs, even a small treatment effect results in a positive expected value of the treatment overall, which explains the prevalence of coupons in e-commerce. However, this cost structure is specific to the focal use case of e-couponing. Coupon targeting is only one possible use case of the proposed decision framework. The framework itself is generic and applies equally in marketing settings with non-zero targeting costs. Higher targeting costs will reduce the fraction of users targeted ceteris paribus.

We now introduce an individual-level targeting policy by estimating the cost of the marketing treatment on the customer level using a response model. Both the two-model and single-model architectures result in a substantial decrease in the fraction of customers treated from 100\% to 72\%--74\%, depending on the estimator. The decrease in the treatment ratio is accompanied by almost no variation in the campaign profit. 
The substantial decrease in the targeting fraction of 26-28\% compared to universal treatment is the direct result of controlling the expected treatment cost for each customer.
The observed positive impact generalizes to customer-level targeting based on the CATE under treatment effect heterogeneity. A hypothetical targeting policy based on the true ITE and the average cost results in a campaign profit of \EUR{60,125}. We find that customer-level estimates of the treatment cost decrease the targeting ratio 
and increase campaign profit. 


In summary, Table \ref{tab:conversion_profit} confirms that considering the probability of customers to accept a costly marketing incentive directly increases campaign profit. This conclusion applies to campaigns considering heterogeneous treatment effects and population-level estimates of the average treatment effect and emphasizes that estimating the expected individual cost based on $p(C_i | X_i=x, T_i=1)$ is practically relevant for customer targeting. For example, in contrast to prior work \cite[e.g.]{gubela2019ConversionUpliftEcommerce}, our analysis implies that targeting customers with a positive response to treatment can be inefficient due to a high conversion probability after treatment and the associated higher expected treatment cost.

\subsection{Profit Implications of Causal Hurdle Models}\label{sec:causal_hurdle_models}
This section complements the analysis of Table \ref{tab:conversion_profit}, in which we have fixed the CATE model to emphasize the link between response probabilities, targeting costs, and campaign profit. We now take the opposite approach and examine the profit impact of CATE estimation. To that end, we assume a constant targeting cost, which we derive from the population average conversion probability. Table \ref{tab:cate_model} reports the results of different instances of the analytical targeting policy and a baseline, which we compare in terms of campaign profit, the fraction of targeted customers (FT), and, to emphasize the quality CATE estimates, the root-mean-squared error (RMSE) of the CATE estimates compared to the simulated ITE on profit. We also consider the TOL on the observed outcomes, which, unlike 
the RMSE, is a feasible metric when the true treatment effect is not simulated and therefore unknown \cite{hitsch2018HeterogeneousTreatmentEffects}.
To put observed results into context, the ATE estimate provides the baseline obtained by a constant estimator, while the oracle ITE guarantees the best results that can be obtained on the data. Kernel density plots showing the distributional fit of the CATE estimates are available in Figure \ref{fig:CATE_densities} in Appendix \ref{sec:additional_results}.

Table \ref{tab:cate_model} confirms the effectiveness of DR and XBCF, which are specifically designed for CATE estimation. Their errors are lower than those of the single- and two-model approach, which agrees with prior comparisons of alternative CATE estimators \cite{kunzel2019MetalearnersEstimatingHeterogeneous, Knaus_causal_ML_MC_study}. Especially XBCF provides the best results in terms of the RMSE and TOL. Notably, we observe a high rank correlation of \textit{Spearman's} $\rho=0.83$ between the RMSE and the TOL. Given that the RMSE cannot be calculated with observational data, high correlation evidences the suitability of using TOL for hyperparameter tuning. 

\begin{table}[htb]
\centering
\caption{CATE-based targeting under population average cost estimates}
\label{tab:cate_model}
{\begin{tabular}{llllrrrr}
\toprule
           &   \multicolumn{3}{c}{Architecture}  & \multicolumn{2}{c}{CATE}  & \multicolumn{2}{c}{Campaign} \\
                     \cmidrule{2-4}\cmidrule(lr){5-6}\cmidrule(lr){7-8}
Policy      & Stages     & CATE Model    & Estimator & RMSE & TOL     & Profit  & FT         \\
\midrule
Baseline    &      -     &        -      &      -    & - &  - &  50,668 &  0.00 \\
\addlinespace[1ex] 
Analytical  &     -      &   ATE         &     -     & 3.77 & 4186.3       &  55,055 &  1.00 \\
\addlinespace[1ex] 
Analytical  &   Hurdle   &  Single-Model &       GBT & 3.02  & 4178.8 &  57,186 &  0.49  \\
Analytical  &   Hurdle   &     Two-Model &       GBT & 3.39  & 4181.7 &  58,335 &  0.52 \\
\addlinespace[1ex] 
Analytical  &  One-Stage &  Single-Model &       GBT & 2.96  & 4180.1 &  57,634 &  0.51\\
Analytical  &  One-Stage &     Two-Model &       GBT & 3.37  & 4180.9 &  57,737 &  0.53 \\
Analytical  &  One-Stage &            DR &       GBT & 2.88  & 4179.7 &  58,022 &  0.54 \\
Analytical  &  One-Stage &          XBCF &      BART & 2.70  & 4177.5 &  57,626 &  0.51 \\
\addlinespace[1ex] 
Analytical  &     -      &        Oracle  &      -   & 0.00  & 4164.5 & 60,125 &   0.56 \\
\bottomrule
\multicolumn{8}{l}{\small FT: fraction of customers targeted.}\\
\end{tabular}}
\end{table}

Considering the single- and two-model hurdle models, we find the former to estimate the CATE more accurately, even outperforming DR in terms of the TOL. This result is notable because a comparison in RMSE and TOL disadvantages the causal hurdle models. Error values are driven by model estimates of the outcome (profit) $Y_i$. Causal hurdle models decompose this quantity according to $Y_i=C_i \cdot V_i$ and estimate conversion probability and value using separate models. It is plausible that a direct estimation of $Y_i$, as implemented in the one-stage approaches, gives a lower error. Overall, the RMSE and TOL results may be taken as evidence that a specialized approach such as XBCF is most suitable if CATE estimation is the main concern. However, campaign profit is arguably the key performance measures for targeting models and Table~\ref{tab:cate_model} indicates that precise CATE estimation and profitable targeting are not equivalent. For example, the two-model hurdle model emerges as most profitable despite exhibiting a  large RMSE. The rank correlation between profit and the RMSE (TOL) also decreases to 0.43 (0.77). However, an overarching finding from Table~\ref{tab:cate_model} is all causal targeting models provide sizeable increases of campaign profit of at least \EUR{2,000}, which equates to a four percent profit increase when compared to the baseline policies. Comparing to the maximum profit of the oracle model further confirms the appealing performance of the ML-based CATE estimators and suggests that they recommend suitable campaign sizes.

\subsection{Profit Implications of the Proposed Analytical Targeting Policy}\label{sec:analytical_policy}
Prior sections evaluate estimates of the CATE and conversion separately. We now examine how the interaction of the treatment and conversion models in the proposed targeting policy impacts campaign profit. Recall that the single- and two-model hurdle models provide an estimate of the conversion probability by design. CATE models that estimate the treatment effect on the profit directly require a separate classification model to predict the conversion rate under treatment.
Table \ref{tab:profit_treatment} compares the proposed analytical targeting policy (Eq. \ref{eq:kappa=0}) to an \textit{empirical} policy that determines the profit-optimal threshold on the CATE estimates through numeric optimization of the overall campaign profit on the training data. 

\begin{table}[htb]
\centering
\caption{Campaign profit for CATE-based targeting under model-based cost estimation}
\label{tab:profit_treatment}
{\begin{tabular}{lllllrr}
\toprule
          &              \multicolumn{4}{c}{Architecture}    & \multicolumn{2}{c}{Campaign} \\
          \cmidrule(lr){2-5}\cmidrule(lr){6-7}
Policy & Stages & CATE Model & Conversion Model  & Estimator & Profit & FT \\
\cmidrule(lr){1-7}
Analytical &   Hurdle   &  Single-Model &   -           &   GBT             &  60,152 &          0.65 \\
Analytical &   Hurdle   &     Two-Model &   -           &   GBT             &  58,967 &          0.66  \\
\addlinespace[1ex] 
Analytical &  One-Stage &  Single-Model &   Separate    &   GBT             &  57,851 &          0.63 \\
Analytical &  One-Stage &     Two-Model &   Separate    &   GBT             &  58,542 &          0.63 \\
Analytical &  One-Stage &            DR &   Separate    &   GBT             &  59,117 &          0.67  \\
Analytical &  One-Stage &          XBCF &   Separate    &   BART/GBT        &  59,069 &          0.61 \\
\cmidrule(lr){1-7}
Empirical &   Hurdle    &  Single-Model &   -           &   GBT             &  57,177 &          0.63 \\
Empirical &   Hurdle    &     Two-Model &   -           &   GBT             &  58,121 &          0.66 \\
\addlinespace[2ex] 
Empirical &  One-Stage  &  Single-Model &   -           &   GBT             &   58,051 &          0.56 \\
Empirical &  One-Stage  &     Two-Model &   -           &   GBT             &  57,535 &          0.68 \\
Empirical &  One-Stage  &            DR &   -           &   GBT             &  57,414 &          0.57 \\
Empirical &  One-Stage  &          XBCF &   -           &   BART            &  57,432 &          0.55 \\

\bottomrule
\multicolumn{7}{l}{\small FT: fractions of customers targeted.}
\end{tabular}}
\end{table}

Table~\ref{tab:profit_treatment} provides strong evidence in favor of our analytical targeting policy. It reports results for several instances of the targeting policies, which vary across modeling stages, CATE and conversion models, and estimators. In every pairwise comparison, the instance of the analytical policy yields higher campaign profit than the corresponding instance of the empirical policy, and only the two strongest instances of the empirical policy give higher profit than the weakest instance of the analytical policy (one-stage single-model approach). The review of related work review has identified the empirical policy as the status-quo in the literature. Recent work \cite{lemmens2020ManagingChurnMaximize} also advocates an empirical targeting policy, which re-emphasizes the competitiveness of this benchmark. However, Table \ref{tab:profit_treatment} facilitates concluding that the proposed analytical policy is substantially better for targeting (coupon) campaigns to the right customers. This is also apparent from the average profits (across all instances), which amount to \EUR{58,950} and \EUR{57,621} for the analytical and empirical policy, respectively, and show the former to raise profit by two percent.

Table~\ref{tab:profit_treatment} suggests that the success of the analytical policy comes from targeting more customers. The analytical policy determines the targeting threshold according to Eq. \ref{eq:kappa=0}, which advocates augmenting a CATE model with an auxiliary model of customers' response probability. Comparing Table~\ref{tab:profit_treatment} to Table~\ref{tab:cate_model}, we find the need of a dedicated response model confirmed. Campaign profits, as well as targeting fractions, for the analytical policy are consistently higher in Table~\ref{tab:profit_treatment}, which accounts for heterogeneity in customer responses. This extra profit, which amounts to \EUR{1193} on average, is lost under the empirical policy. There, the average profit difference between an approach of Table~\ref{tab:profit_treatment} and Table~\ref{tab:cate_model} amounts to \EUR{-135}, implying that empirically tuning the targeting threshold performs worse than applying the proposed decision rule even if assuming customer response probabilities to be constant.   

Compared to an empirical optimization of the targeting threshold, it is plausible that the targeting rule of Eq. \ref{eq:decision_rule} is more robust if strong heterogeneity in customer response probabilities and/or treatment effects is present. Given that the true ITE is unobservable, it is worth discussing whether the observed results depend on our simulation of the ITE. Specifically, we cannot rule out the possibility that our data exhibits stronger heterogeneity in the ITE compared to how customers actually respond to a coupon offer in practice. Strong heterogeneity suggests high variance among CATE estimates. Deciding on a targeting fraction by optimizing over these estimates might lack reliability. Therefore, we expect strong heterogeneity to benefit the analytical policy. Comparing the analytical and the empirical policy in the field is highly recommended to confirm the superiority of the former, and is a fruitful avenue for future research. We are confident to have used the available data in the best possible way and in agreement with prior literature \cite{lemmens2020ManagingChurnMaximize, Knaus_causal_ML_MC_study, hitsch2018HeterogeneousTreatmentEffects}. Considering that field experiments require a strong commitment of line managers and consume a sizeable amount of resources, we also hope that the results observed here aid the organization of future field studies by evidencing the potential of the analytical policy and, thereby, justifying investments into field experimentation.    

Last, Table~\ref{tab:profit_treatment} facilitates conclusions concerning the relative merit of alternative implementations of a targeting policy. Concentrating on the analytical policy, we note the strong performance of the single-model hurdle model. It provides the highest campaign profit in the comparison, outperforming even the state-of-the-art CATE models DR and XBCF. A feature of the single-model hurdle approach is that its implementation requires only two component models (see Table \ref{tab:experiment_summary}). Compared to the five component models in the one-stage DR approach, the runner-up in the comparison, the single-model hurdle model should be substantially easier to maintain in practice. More generally, Table~\ref{tab:profit_treatment} supports the view that causal hurdle models are a suitable vehicle for the coupon targeting campaign considered here. For both, the analytical and empirical targeting policy, we find the overall best approach to be among the causal hurdle models and their profit figures suggest that these models are at least competitive to alternative means of implementing a targeting policy. We also note that the single- and two-model hurdle models perform better than their one-stage counterparts in three of four cases, which corroborates the previous conclusion. Finally, a distinct advantage of the hurdle structure is that the stage-wise estimates facilitate interpreting the treatment effect on customer conversion and spending behavior separately. These insights benefit marketing practice in that they can inform decisions concerning, for example, the design of a marketing treatment and clarify its effectiveness in increasing customer conversion versus customer value. We exemplify this feature in Appendix \ref{appendix:hurdle_model_features}.

\section{Conclusion}\label{conclusion}
We presented an analysis of the customer targeting decision problem under different types of variable costs and proposed a modeling framework to estimate relevant decision variables. The derived policy recommendations and the framework are generic and guide targeting decisions marketing campaign settings with various cost structures. The empirical analysis of the modeling framework demonstrates that it supports arbitrary approaches for estimating the CATE include recently proposed ML-based techniques. We recommend instantiating the framework by means of a causal hurdle model. 
Its design streamlines modeling building by reusing component models, which marketing analysts can estimate using their preferred ML algorithms. 
We assessed our propositions in a specific campaign setting associated with targeting e-coupons. The observed results confirmed the proposed analytical targeting policy, the necessity to discount response-dependent costs in targeting decisions, the modeling framework, and the effectiveness of causal hurdle models. For the specific campaign under study, these components were shown to increase campaign profit compared to benchmarks when assessed individually and jointly. 

Our results have implications for marketing practice and academia.
While customer targeting based on expected profit has been used to optimize campaigns, prior work does not consider marketing incentives that are conditioned on customer responses. We identify these common marketing incentives as a type of stochastic variable cost. Our analysis of the targeting decision problem under customer response-dependent costs shows that estimating the expected cost requires an estimate of the customer response conditional on treatment. A central implication for the customer targeting literature is that profit-maximizing targeting requires modeling the effect of the marketing treatment and the net customer response under treatment. 

For marketing practice, the paper raises awareness for the stochastic nature of response-dependent costs. Many campaigns, especially in digital marketing, involve incentives with low targeting-dependent and high response-dependent variable costs. Full recognition of the latter is crucial if they are the main driver of treatment costs. To our best knowledge, the corresponding policy recommendations and modeling methodologies were lacking and are originally provided here. Targeting based on uplift models, which has only recently gained wider dissemination \cite{devriendt2019WhyYouShould}, disregards response-dependent costs. Our decision analysis predicts such an approach to provide sub-optimal decisions. The observed results confirm this prediction and demonstrate, for the specific campaign under study, the recognition of response-dependent costs in a targeting policy to substantially raise profit. 
Consequently, we recommend revisiting deployed targeting practices and checking the degree to which these account for relevant treatment cost components. The paper systematizes variable cost components for different types of campaigns and identifies the decision parameters that govern optimal targeting. Levering established ML algorithms and prescribing their use for estimating these parameters, the proposed modeling framework offers marketers an actionable approach to enhance campaign planning practices.   


\bibliographystyle{unsrt}
\bibliography{references.bib}


\newpage
\appendix

\section{Relation to Previous Formulations of Churn Campaign Profit}\label{sec:churn_profit_formula}
A popular definition of the profit of a customer retention campaign \cite{verbeke2012NewInsightsChurn,devriendt2019WhyYouShould} is given by \cite{neslin2006DefectionDetectionMeasuring}:
\[
\Pi = N \alpha  \left[  \beta\gamma(V - \delta - \kappa) + \beta(1-\gamma)(-\kappa) + (1-\beta)(-\delta-\kappa)  \right] - A
\]

with\\
N: Number of customers\\
\(\alpha\): Ratio of customers targeted\\
\(V\): The value of the customer to the company, $CLV$ in their original notation\\
\(\beta\): Fraction of (targeted) customers who would churn\\
\(\gamma\): Fraction of (targeted) customers who decide to remain when
receiving the marketing incentive\\
\(\delta\): The cost of the marketing incentive if it is accepted\\
\(\kappa\): The cost of contacting the customer with the marketing
incentive\\
\(A\): The fixed cost of running the retention campaign

    The number of customers targeted by the campaign and the fixed costs are relevant to calculate the overall campaign profit, but do not affect the targeting decision for a single customer. The profit estimate relevant for customer targeting is thus the part in square brackets:
%
\[
\pi_i = \beta_i \gamma_i(V_i \textcolor{blue}{-\delta}    
) + \beta_i(1-\gamma_i)(\textcolor{red}{-\kappa}) + (1-\beta_i)(\textcolor{blue}{-\delta }\textcolor{red}{-\kappa})
\]

    We will show that this expression is equivalent to the proposed decision policy (Eq. \ref{eq:decision_rule}) under restrictive assumptions. 
%
    Using the additive property of the probabilities \(\beta_i\) and \((1-\beta_i)\) and \(\gamma_i\) and \((1-\gamma_i)\), we can summarize the terms:
    \begin{align*}
\pi_i &= \beta_i \gamma_i(V_i) + 
\textcolor{blue}{\beta_i\gamma_i (-\delta) + (1-\beta_i) (-\delta)} \textcolor{red}{+\beta_i (-\kappa) + (1-\beta_i)(-\kappa)}  \\
       &= \beta_i \gamma_i V_i + \textcolor{blue}{\beta_i\gamma_i (-\delta) + (1-\beta_i)-\delta} \textcolor{red}{-\kappa}\\
       &= \beta_i \gamma_i V_i  - \textcolor{blue}{\delta(\beta_i\gamma_i + 1 -\beta_i)} \textcolor{red}{-\kappa}\\
       &= \beta_i \gamma_i V_i - \textcolor{blue}{\delta(1-\beta_i(1-\gamma_i)) } \textcolor{red}{-\kappa}
\end{align*}

    We will target a customer if the profit is positive, i.e.
\begin{align}
\label{eq:churn_campaign_neslin}
\beta_i \gamma_i V_i - (1-\beta_i(1-\gamma_i)) \delta - \kappa > 0
\end{align}

    In Eq. \ref{eq:decision_rule}, we propose the decision rule
\[
p_i(1) (V_i(1) - \delta) - \kappa > p_i(0) \cdot V_i(0)
\]

Assuming that the value of the customer is not influenced by the marketing incentive $V(1)=V(0)=V$ allows us the rearrange the inequality to 
\begin{align}
\label{eq:churn_campaign_ours}
(p_i(1) - p_i(0))V_i - p_i(1) \delta - \kappa > 0
\end{align}

Eq. \ref{eq:churn_campaign_neslin} and Eq. \ref{eq:churn_campaign_ours} are equivalent if the following equalities hold:
\begin{align*}
p_i(1) &= (1-\beta_i(1-\gamma_i)) \\
p_i(1)-p_i(0) &= \beta_i\gamma_i   \\         
\end{align*}
  In words, we require $p_i(1)$ to be the complement to the probability for a customer to plan to churn and churn even when offered the treatment. The complimentary event is for a customer not to plan to churn or to plan to churn but remain after treatment; or simply, the probability of the customer to stay when given treatment $p_i(1)$.
  
  We further require $p_i(1)-p_i(0)$ to be the probability of a customer to plan to churn and to not churn when offered the treatment. As $\beta_i \cdot \gamma_i \in [0;1]$, this equality holds under the assumption that the treatment effect is strictly positive, i.e. ${p_i(1)-p_i(0) \in [0;1]}$. However, we know that the treatment effect on the response probability, \(p_i(1)-p_i(0)\), is in principle bounded in $[-1,1]$ and that negative effects are a critical issue in churn campaigns in practice \cite{ascarza2018RetentionFutilityTargeting}. 
Under the previous campaign profit formulation, we see that \(\beta_i\gamma_i=0\) if either \(\beta_i\) or \(\gamma_i\) or both are zero. In words, the campaign has no effect if no customers consider to churn or no customers accept the marketing incentive when offered. This conflicts with the observation that when no customers plan to churn, the campaign may have a net negative effect by priming inattentive customers to churn. Specifically, the shortcoming of the customer profit proposed by \cite{neslin2006DefectionDetectionMeasuring} is that it implicitly assumes a positive treatment effect by restricting the action space of the customer to $\gamma \in \{\text{Accept treatment}, \text{Disregard treatment}\}$. 

The customer's action space when treatment may have a negative effect is better represented if we include a third action, $\gamma \in \{\text{Accept treatment}, \text{Disregard treatment}, \text{Reject treatment}\}$. Negative reaction to treatment is of particular concern in retention campaigns, where the campaign can act as a churn trigger for otherwise passive customers.
We can derive the campaign profit under the assumption of the extended action space to show that the resulting profit function corresponds to the conclusions derived from the potential outcome framework. Denote the probability that a customer will react negatively to the offer by $\lambda$. The customer will accept the offer with probability $\gamma$ and will not accept the offer with a probability $1-\gamma-\lambda$.
Under the previous assumption that customers who do not plan to churn will not reject the offer, it follows that $(1-\beta)(1-\gamma-\lambda)=0$. That assumption is not strictly necessary for the analysis but relates more closely to the analysis in \cite{neslin2006DefectionDetectionMeasuring}.

The enhanced churn campaign profit formula is (with changes in red)
$$
\Pi_i = \beta\gamma(CLV - \delta -\kappa) + {\color{red}{(1-\beta)\lambda(-CLV-\kappa)}} + \beta(1-\gamma{\color{red}{-\lambda}})(-\kappa) + (1-\beta)(-\delta-\kappa)  
$$
We see that the extended formulation of the campaign profit is equal to the profit under response-dependent cost (Eq. \ref{eq:decision_rule}) when the following equalities hold
\begin{align*}
p_i(1)-p_i(0) &= \beta\gamma - (1-\beta)\lambda   \\         
p_i(1) &= (1-\beta(1-\gamma)), 
\quad \text{and following from these}    \\
p(0) &= 1-\beta.
\end{align*}
We see that the potential outcomes in the proposed formulation of profit under response-dependent cost have an intuitive interpretation for churn campaigns. $p_i(0)$ is the probability of a customer to make the plan to churn. $p_i(1)$ is the complementary probability of a customer to make a plan to churn and churn even when offered the treatment. That includes customers who make no plan to churn and those that are convinced by the incentive. $p_i(1)-p_i(0)$ is the probability of a customer to make the plan to churn and not churn when offered the treatment minus the probability of the customer to have no plan to churn but to churn when offered the treatment. 

This interpretation covers empirical cases of campaign performance that are not covered by the original formulation. In particular, a campaign has no effect if A) all customer consider to churn but nobody reacts to the incentive, B) no customers consider to churn and the campaign has no negative effects or C) if the number of customers who consider churning and are convinced by the campaign and the number of customers who were not planning to churn but are negatively affected by the campaign is exactly equal. This last case of where some churn is prevented but other customers are triggered is not covered by the original formulation. Formally, we observe $\beta\gamma - (1-\beta)\lambda=0$ if A) $\beta=1$ and $\gamma=0$, B) $\beta=0$ and $\lambda=0$ or C) $\beta\gamma = (1-\beta)\lambda$. 

We conclude that the proposed decision framework is a generalization of \cite{neslin2006DefectionDetectionMeasuring}'s campaign profit function to cases where a customer may react adversely to the treatment. As an alternative formulation to calculate the overall churn campaign profit, we propose for the general case with negative treatment effects:
\[
\Pi = N \alpha  \left[ \beta\gamma(V - \delta -\kappa) + (1-\beta)\lambda(-V-\kappa) + \beta(1-\gamma -\lambda)(-\kappa) + (1-\beta)(-\delta-\kappa) \right] -A
\]
Only in cases with no or little variation in customer sensitivity to the marketing treatment and a constant customer lifetime value can the churn campaign profit be simplified to:
\[
\Pi = N\alpha \left[ \hat{\tau}_{ATE} V_i - p_i(1)\delta -\kappa\right]  - A
\]

\section{Treatment Effect Simulation}\label{appendix:te_simulation}
The main body of the paper summarizes our approach to simulate treatment effects based on the observed data. To further improve clarity, we elaborate on the simulation approach in this section. 

To evaluate the proposed targeting rule and modeling framework, we simulate the treatment effect on customer spending on the actual customer data in an empirical Monte Carlo study. Simulation is suitable because implementing the profit-maximizing policy involves estimating the CATE of a coupon, the true value of which cannot be observed in practice.
Specifically, we simulate a treatment effect as ground truth for evaluation while using as much actual data as possible. This avoids oversimplifying the data generating process by assuming simple relations among the covariates and between the covariates and the outcome \cite{nie2017QuasiOracleEstimationHeterogeneous, Knaus_causal_ML_MC_study}. In addition, we strive to simulate realistically complex customer responses to the marketing incentive by a process that depends nonlinearly on a subset of the observed covariates and draw on recent work of \cite{hitsch2018HeterogeneousTreatmentEffects} to define the range of the simulated ITE on response probability and value. Thus, the goal of our empirical Monte Carlo approach is to evaluate the proposed targeting framework on realistic customer data while controlling the ITE to facilitate evaluation in comparison to the ground truth.  

We simulate the treatment effect dependent on the covariates for each user and allow for heterogeneity in the effect on customer purchase probability and purchase value. To this end, we define the overall treatment effect $\tau(X_{\tau})$ as a combination of the treatment effect on the conversion probability $\tau_C(X_{\tau})$ and the purchase value conditional on a purchase $\tau_V(X_{\tau})$, where $X_\tau$ is a subset of $k=11$ covariates describing the customer's entry channel and the outcome of previous visits, if any. We draw each treatment effect as a nonlinear combination of covariates 
characterized by two weight matrices whose elements are drawn randomly from independent Gaussian distributions
\begin{align*}
W_{C,1}, W_{C,2}, W_{V,1}, W_{V,2} &\sim \mathcal{N}(0_k, I_k)\\
\tau_C(X) &= \sigma(X_\tau^\top W_{C,1})^{\top}W_{C,2} \\
\tau_V(X) &= \sigma(X_\tau^\top W_{V,1})^{\top}W_{V,2} 
\end{align*}
where $\sigma$ is the logistic function. 
Both treatment effects are centered and scaled. We scale the ITE distribution to have most of its mass in the range [0;10] and a positive average effect. Based on previous work, we consider these values representative for direct marketing applications \cite[Figure 12]{hitsch2018HeterogeneousTreatmentEffects}. For the ITE on response probability, we center the distribution around an ATE of 5 percentage points and truncate the simulated values to the range [-0.1,0.15]. For the ITE on response value, we center the distribution around an ATE of \EUR{1} and truncate the simulated values to the range [-10,10].
The treatment effect is thus generated by a process that depends non-linearly on a subset of the observed covariates and their interactions. We argue that this approach simulates realistically complex responses of customers to the marketing incentive. 

To preserve the complexity of the customer purchase process, we add the simulated treatment effects to a non-parametric estimate of the purchase probability and value. Note that the purchase probability for each users is unknown because we only observe conversion as a binary event and that the potential purchase value for users who have not completed a purchase is unknown because we observe the final checkout value of a purchase after the purchase is completed. We estimate these quantities by training two weakly regularized GBT ensemble models on the covariates to approximate the unknown relation between the covariates and each outcome.
A gradient boosting classification tree ensembles estimates the underlying conversion probability for all users and a regression tree ensemble trained on users who have made a purchase estimates the purchase value for the users who have not. Each customer is then assigned to the treatment or control condition randomly with probability 0.5. For treatment group customers, we add the simulated treatment effects on purchase probability and value to the estimate. Lastly, the observed purchase event for each customer is drawn at random from a Bernoulli distribution with the customer's purchase probability and the purchase value set to the potential value when a purchase takes place and to zero otherwise. The overall treatment effect on user spending is the difference between the spending of the user if assigned to the treatment and the spending if assigned to the control condition. 

\section{Treatment Effect Decomposition in Hurdle Models}\label{appendix:hurdle_model_features}
\begin{figure}[tb]
  \centering
  \small
    \captionsetup{width=0.95\textwidth}
  \begin{subfigure}[h]{0.48\textwidth}
\centering
\captionsetup{justification=centering}
    \includegraphics[height=5.8cm]{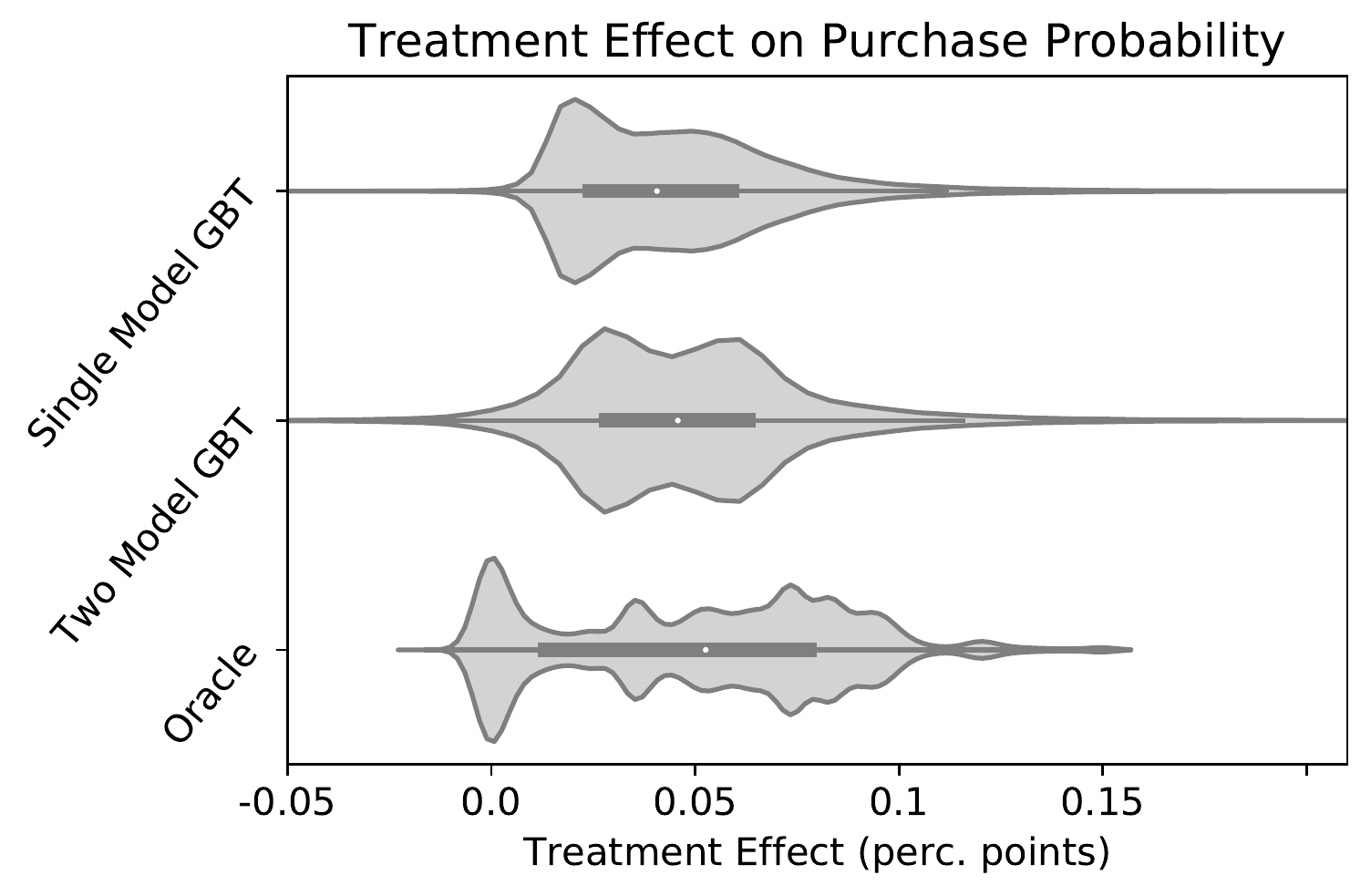}
  \end{subfigure}
\hspace{1em}%
     \begin{subfigure}[h]{0.48\textwidth}
  \centering
  \captionsetup{justification=centering}
    \includegraphics[clip=true,trim=2.9cm 0 0 0,height=5.8cm]{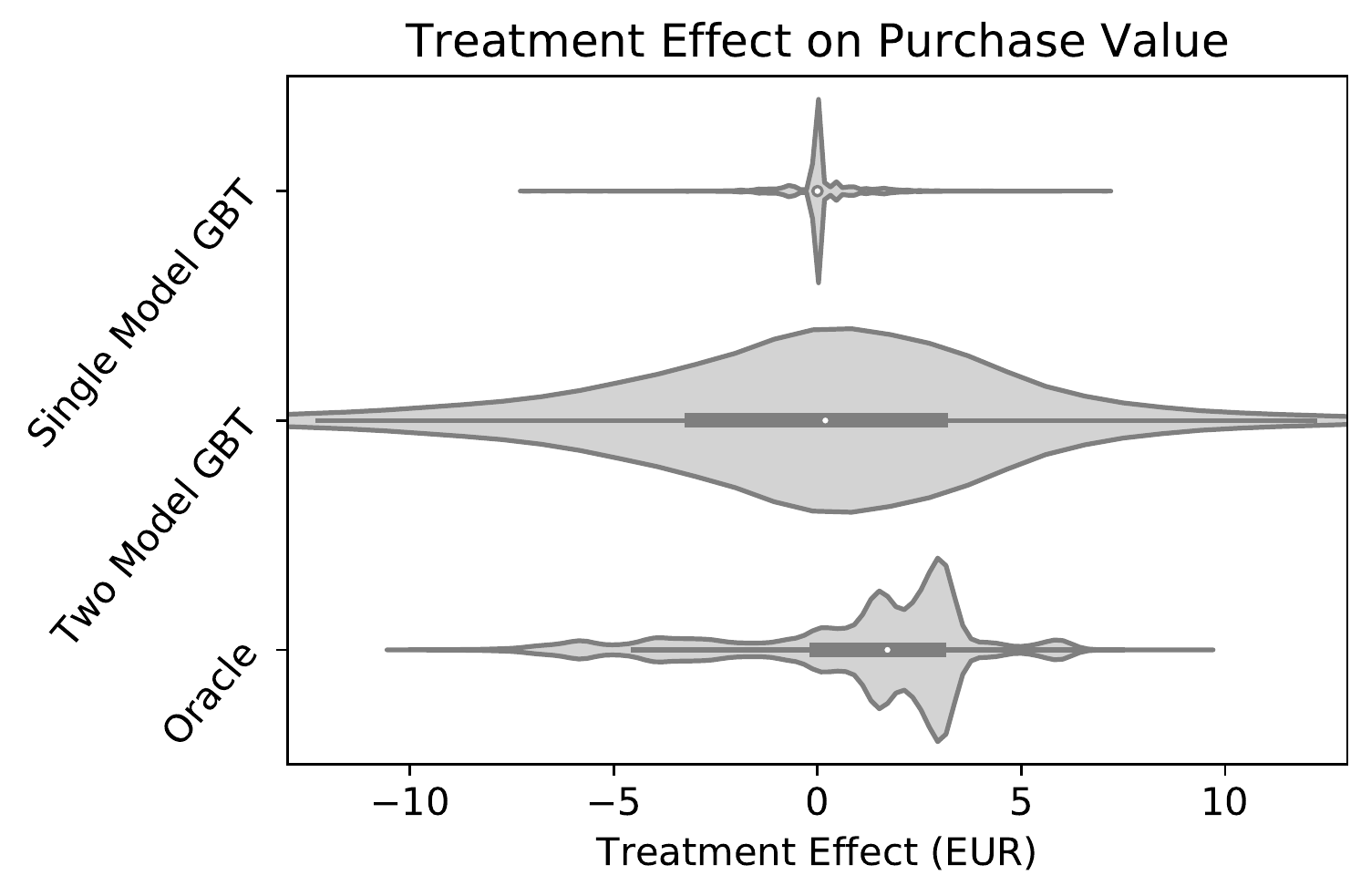}
  \end{subfigure}
  \caption{Treatment effect on the two stages of the customer purchase process. The two-stage structure of hurdle models estimates the overall treatment effect on spending as a combination of the effect on purchase probability and purchase value. The violin plots are scaled to the same width for better comparability. Each plot includes a boxplot, which indicates the median and interquartile-range.}
     \label{fig:hurdle_treatment_distributions}
\end{figure}

Hurdle models are only one option to implement the proposed analytical targeting policy. However, a distinct advantage of the hurdle structure is that the stage-wise estimates facilitate interpreting the treatment effect on customer conversion and spending behavior separately. These insights benefit marketing practice in that they can inform decisions concerning, for example, the design of a marketing treatment and clarify its effectiveness in increasing customer conversion versus customer value. To exemplify this feature, Figure \ref{fig:hurdle_treatment_distributions} shows the kernel density of oracle prediction and the CATE estimates of the causal hurdle models for the purchase probability and the purchase value. With mean estimates of, respectively, 0.45 and 0.48, the single- and two-model hurdle model estimate the simulated ATE on purchase probability of 5 five percentage point approximately correctly. However, we also observe the models to face difficulties in fully capturing the complex functional form of the simulated CATE. 

For purchase value, a mean estimate of \EUR{0.09} and -\EUR{-0.23} for the single- and two-model hurdle model, respectively, does not approximate the simulated ATE of \EUR{1.00} well. Given that the purchase value in the second stage is estimated from the customers who have completed their purchase, it may be that the data of the 8,452 purchasers, which were available to this study, did not suffice to detect the small simulated treatment effect. Capturing this effect may have been complicated further by the IPW selection bias correction. Incorporating the first-stage purchase probability model, the correction propagates inaccuracies in that model, as observed in the left-hand plot in Figure \ref{tab:conversion_quality}, to the second stage. Despite the flexible GBT model specification, customer behavior is complex and a more informative model specification or more granular customer information than was available to this study may have been necessary to correct for variables determining the purchase decision. The large scale of many digital marketing campaigns could heal these problems and facilitate more robust modeling of conditional outcomes and small treatment effects. 


We conclude that the analysis of the treatment effect at each stage of the hurdle model has potential for marketing applications. Our results show how such an analysis can be conducted in principle but indicate that further research is needed to clarify the limits of treatment effect estimation at this granularity in practice and improve on the proposed hurdle specification.

\section{Additional Evaluation Results}\label{sec:additional_results}

Table \ref{tab:conversion_quality} shows the quality of predictions for the conversion probability conditional on treatment. Recall that the single-model hurdle model includes the treatment indicator as a covariate into the model. The two-model hurdle model estimates four separate models, one of which predicts the conversion probability within the treatment group. Note that the default approach, which separates treatment effect estimation and conversion prediction, also requires the estimation of an identical conversion model. This is the redundancy that the proposed causal hurdle framework avoids. We find no substantial difference in the area-under-the-ROC-curve (ROC-AUC) or the Brier score, which indicates model calibration.

   \begin{table}[htbp]
\centering
\caption{Quality of model estimates for the prediction of conversion under treatment}
\label{tab:conversion_quality}   
   \begin{tabular}{lllrr}
\toprule
\multicolumn{3}{c}{Architecture} \\
\cmidrule{1-3}
    Stages & Specification & Estimator&  ROC-AUC &  Brier Score \\
\midrule
Hurdle/Distinct & Two-Model  & Linear &    0.654 &        0.102 \\
Hurdle/Distinct & Two-Model & GBT &  0.671 &        0.101 \\
\addlinespace[1ex]
Hurdle Model & Single-Model & GBT &   0.670 &        0.100 \\
\bottomrule
\end{tabular}
\end{table}


\begin{figure}[htbp]
\fxwarning{Consider fixing the bandwidth of the kernel to avoid any artifacts in the plots}
\captionsetup{width=0.9\textwidth}
  \centering
  \small
  
    \begin{subfigure}[h]{0.45\textwidth}
\centering
  \captionsetup{justification=centering}
    \includegraphics[width=\textwidth]{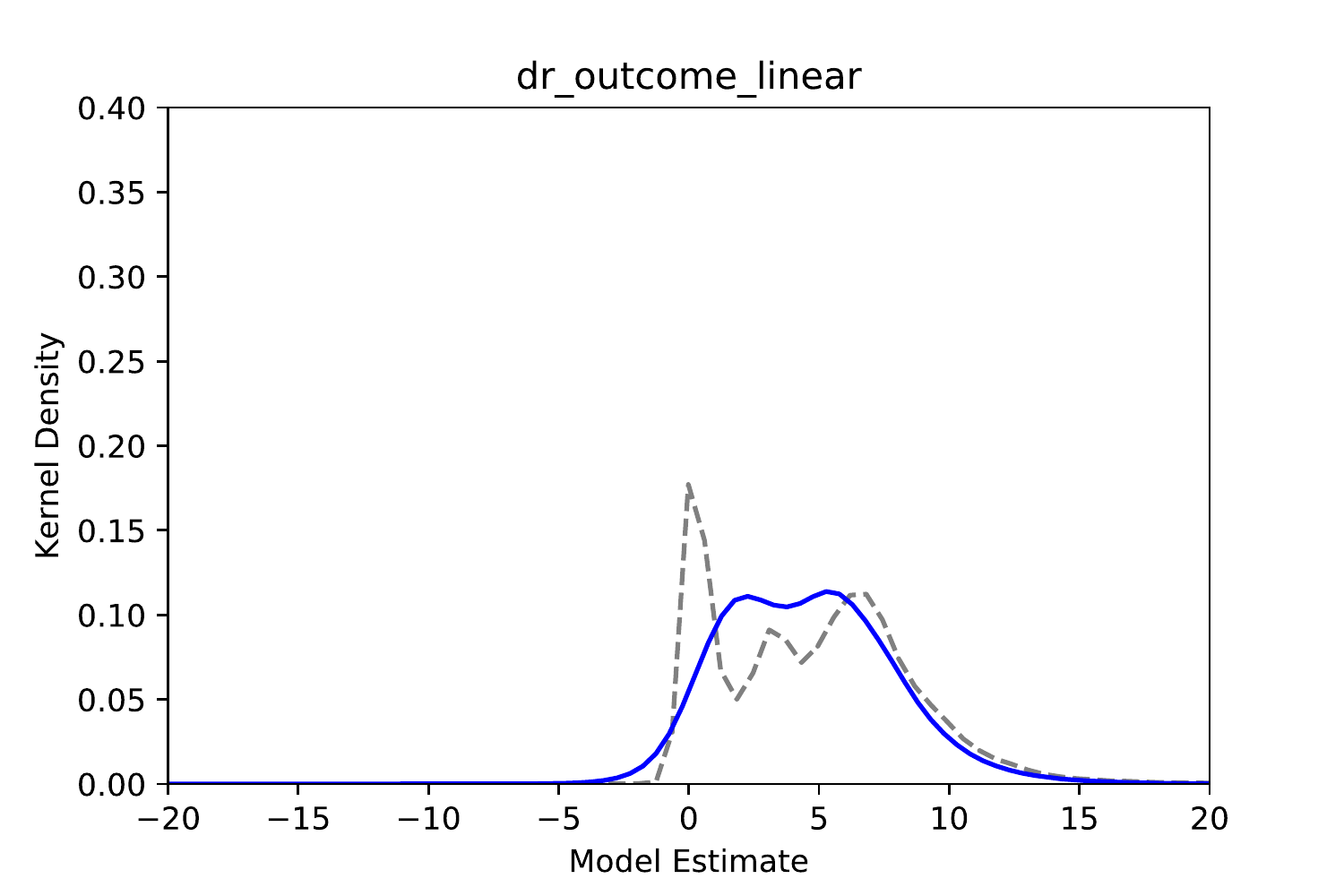}
    \caption{One-Stage DR Linear}
  \end{subfigure}
\hspace{1em}%
     \begin{subfigure}[h]{0.45\textwidth}
  \centering
  \captionsetup{justification=centering}
    \includegraphics[width=\textwidth]{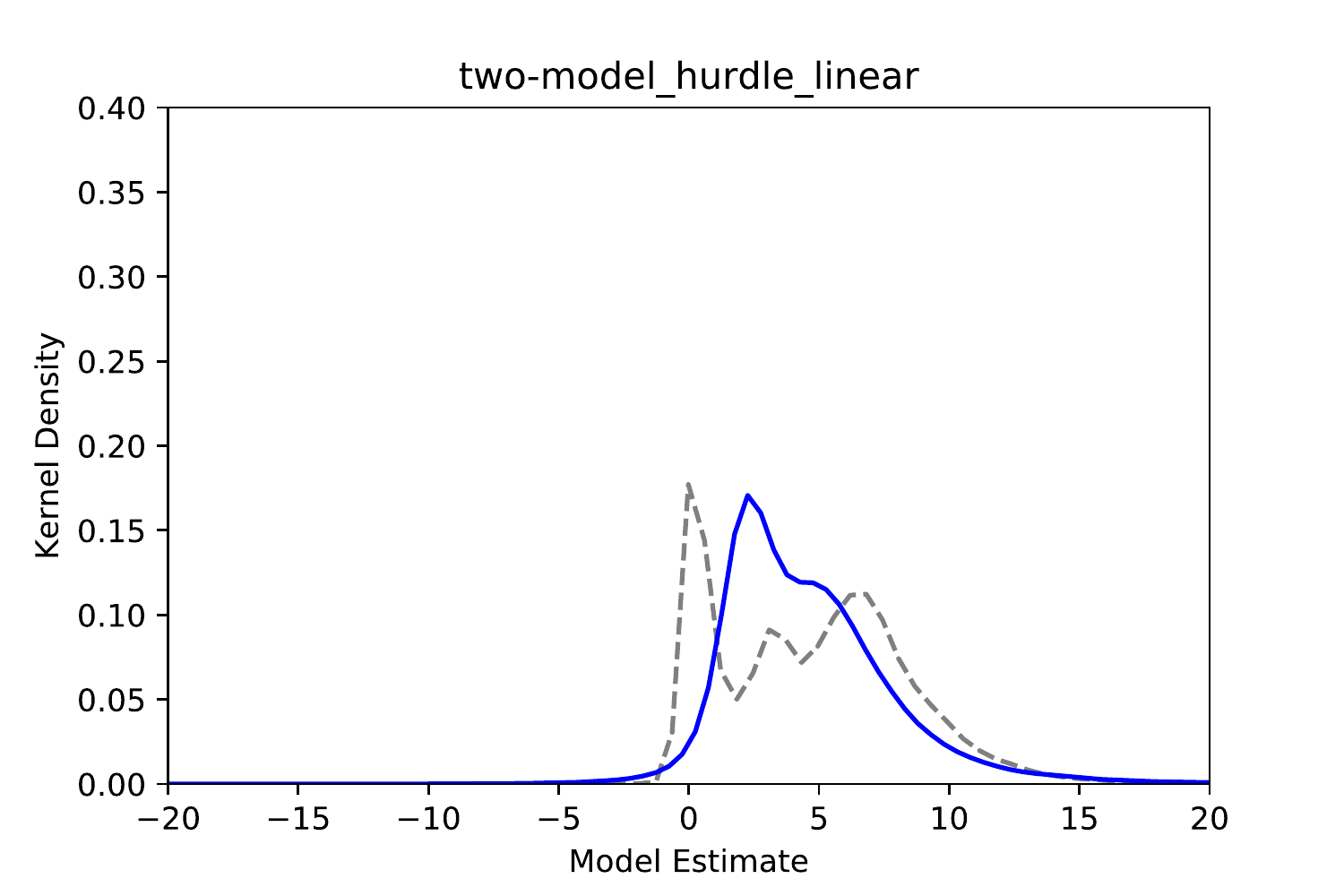}
    \caption{Hurdle Two-Model Linear}
  \end{subfigure}
  
  \begin{subfigure}[h]{0.45\textwidth}
\centering
  \captionsetup{justification=centering}
    \includegraphics[width=\textwidth]{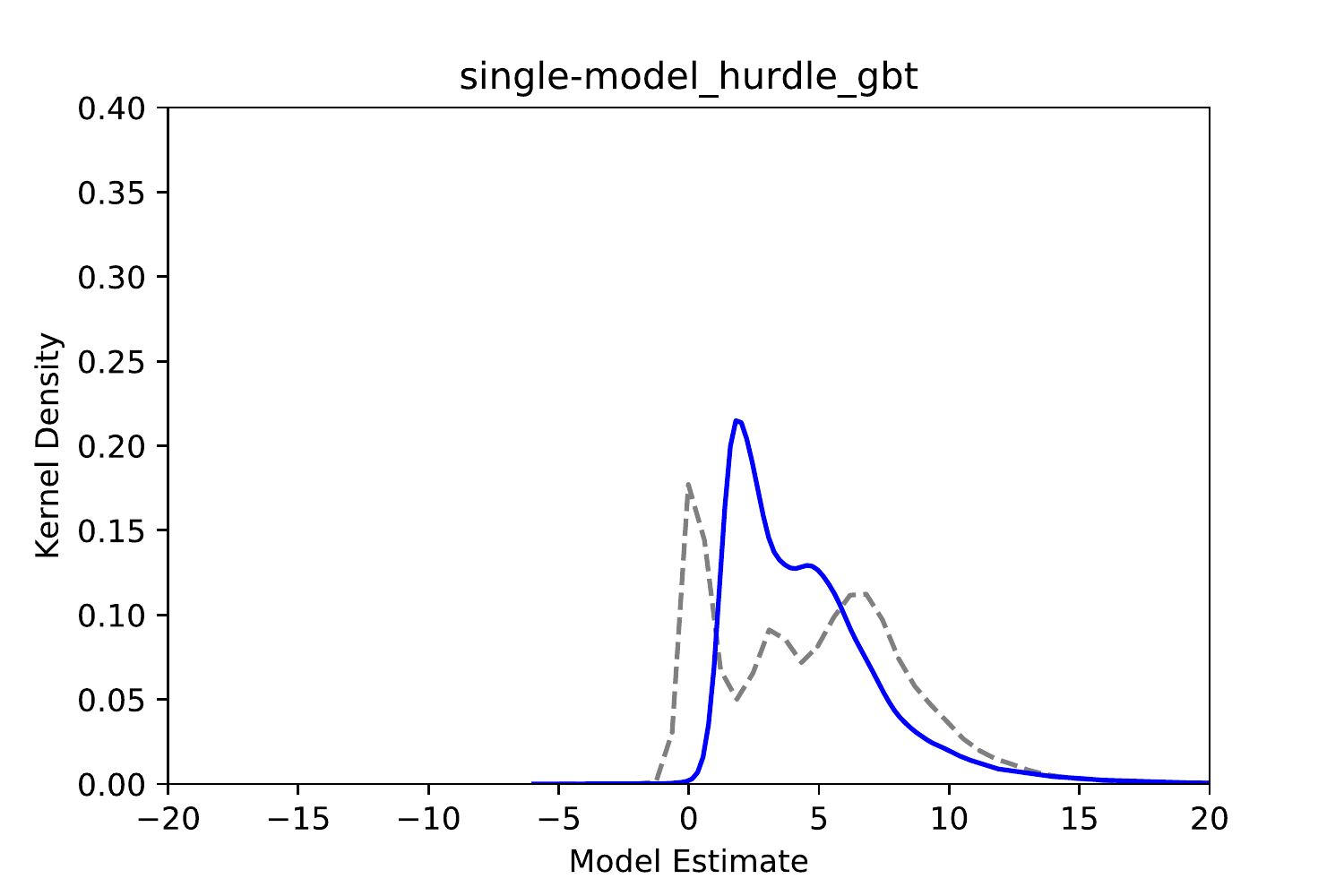}
    \caption{Hurdle Single-Model GBT}
  \end{subfigure}
\hspace{1em}%
     \begin{subfigure}[h]{0.45\textwidth}
  \centering
  \captionsetup{justification=centering}
    \includegraphics[width=\textwidth]{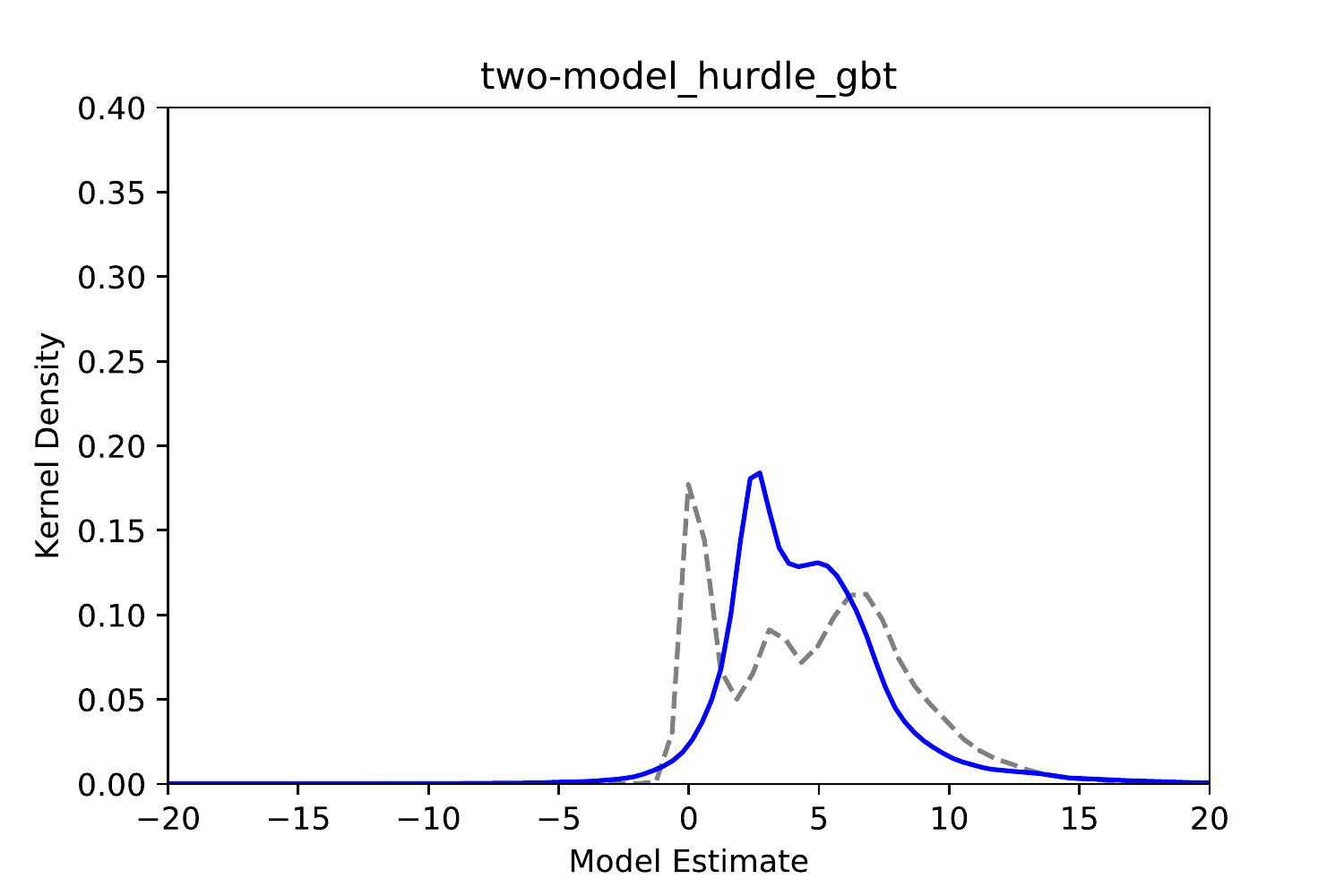}
    \caption{Hurdle Two-Model GBT}
  \end{subfigure}
  
     \begin{subfigure}[h]{0.45\textwidth}
  \centering
  \captionsetup{justification=centering}
    \includegraphics[width=\textwidth]{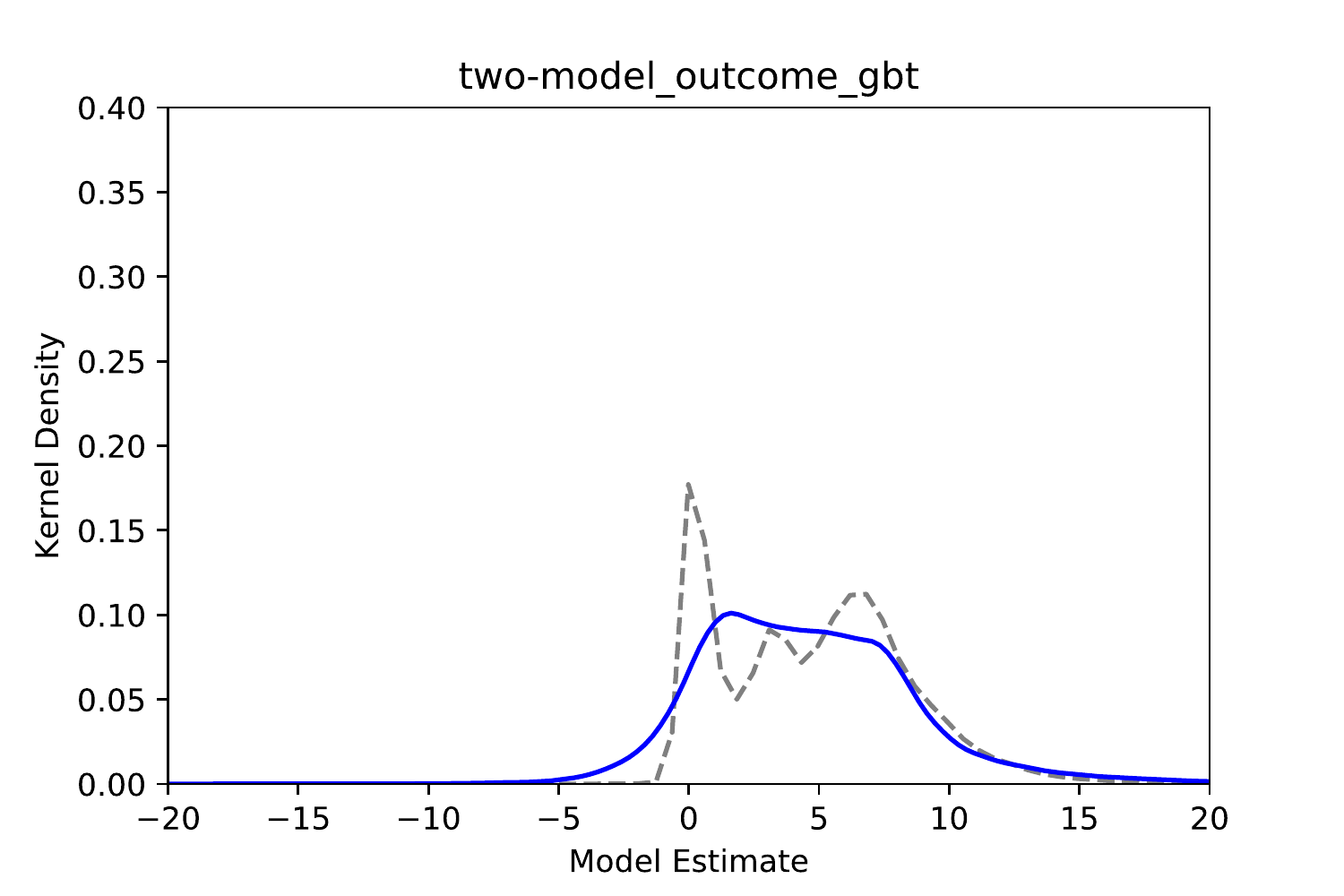}
    \caption{One-Stage Doubly-Robust GBT}
  \end{subfigure}
  \hspace{1em}%
  \begin{subfigure}[h]{0.45\textwidth}
  \centering
  \captionsetup{justification=centering}
    \includegraphics[width=\textwidth]{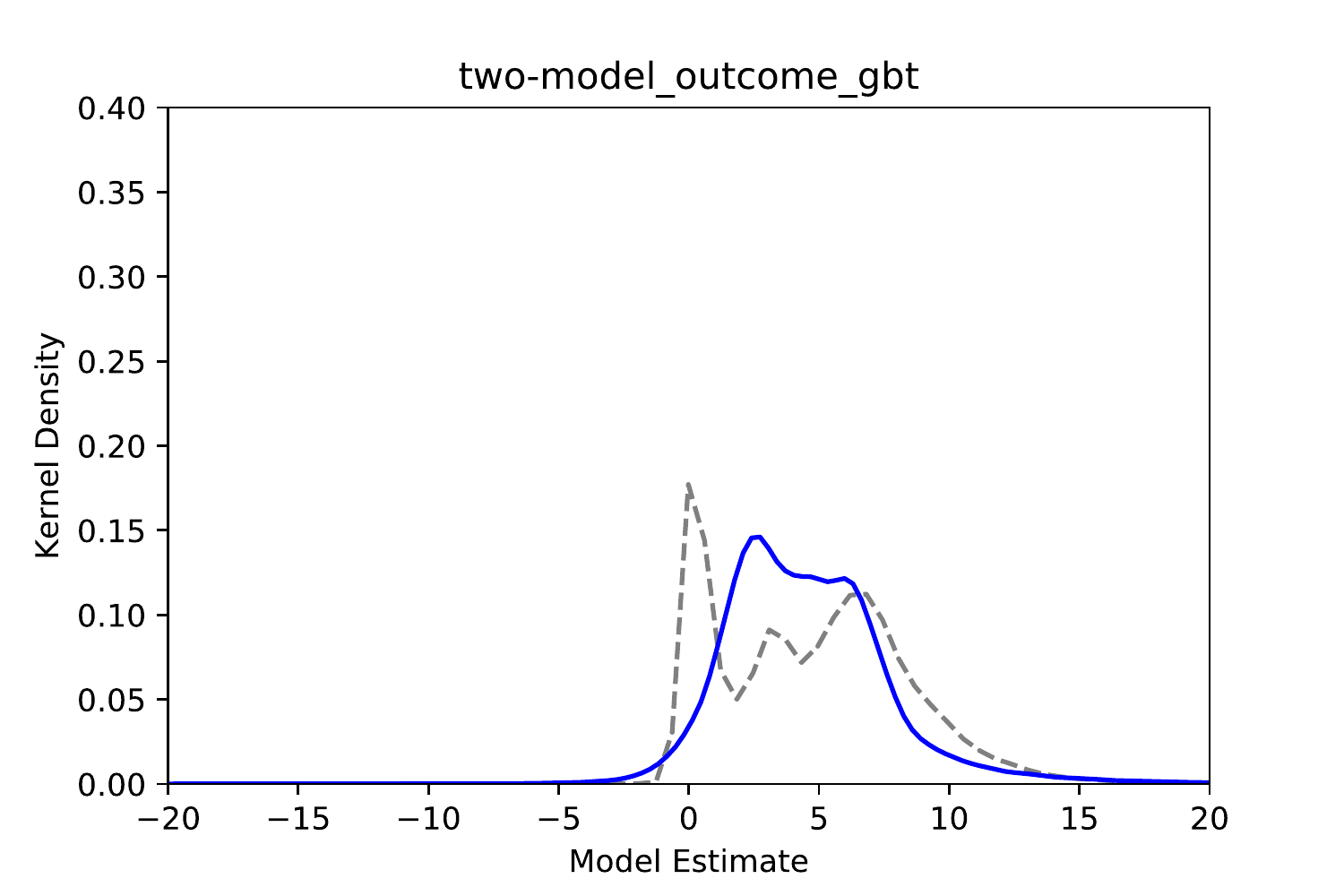}
    \caption{One-Stage Two-Model GBT}
  \end{subfigure}
    \caption{Kernel density plot of the CATE on the outcome as estimated by selected linear model (top), hurdle models (middle) and one-stage models (bottom). The dotted line shows the oracle prediction.} 
  \label{fig:CATE_densities}
\end{figure}

Figure \ref{fig:CATE_densities} depicts the kernel density plot for the treatment estimation approaches and the GBT specification. We combine the out-of-sample estimates for each iteration of the cross-validation procedure to obtain out-of-sample estimates for the full dataset. The dotted line shows the kernel density of the simulated ITE. 

We observe that no approach fully captures the minor mode of the distribution to the left. The hurdle single-model GBT approach in addition shows a slight shift from the major mode of the distribution that relates to the worse precision reported in Table \ref{tab:cate_model}.

The support of the linear model specifications extends beyond the actual range of the simulated treatment effects and beyond the range shown in the figure. For a small set of observations, we observe predicted treatment effects beyond the range [-100;100] that explain the high statistical error reported in Table \ref{tab:cate_model}. For the remaining observations, we observe a reasonable fit of the true treatment effect distribution. The general fit explains the profitability of the linear specification for the targeting policy as observations with weak support, for which linear extrapolation fails, by definition make up only a minority of cases in the data.

\end{document}